\documentclass[prd,aps,twocolumn,a4paper,showpacs,floatfix,nofootinbib]{revtex4}

\usepackage{amsmath,amssymb,amsthm}
\usepackage{wasysym}
\usepackage{mathrsfs}
\usepackage{multirow}
\usepackage{graphicx}
\usepackage{psfrag}
\usepackage{epstopdf}
\usepackage{color}

\def\e{{\rm e}}

\def\Amax{A^{\rm max}_{\ell m}}
\def\lm{{\ell m}}

\def\ii{{\rm i}}

\def\lm{{\ell m}}

\def\hk{{\hat{\hat k}}}

\def\I{{\cal I}}

\def\R{{\cal R}}

\def\ie{{i.e.}~}

\newcommand{\be}{\begin{equation}}  
\newcommand{\ee}{\end{equation}}
\newcommand{\bea}{\begin{eqnarray}}           
\newcommand{\eea}{\end{eqnarray}} 
\newcommand{\beqn}{\begin{eqnarray*}}
\newcommand{\eeqn}{\end{eqnarray*}}
\newcommand{\ba}{\begin{align}}
\newcommand{\ea}{\end{align}}

\begin{document}

\title{Binary black hole coalescence in the extreme-mass-ratio limit:\\
  testing and improving the effective-one-body multipolar waveform}

\author{Sebastiano \surname{Bernuzzi}$^1$}
\author{Alessandro \surname{Nagar}$^2$}
\author{An{\i}l \surname{Zengino$\mathrm{\breve{g}}$lu}$^3$} 

%
\affiliation{$^1$Theoretical Physics Institute, University of Jena,
  07743 Jena, Germany}

%
\affiliation{$^2$Institut des Hautes Etudes Scientifiques, 91440
  Bures-sur-Yvette, France}
 
%
\affiliation{$^3$Theoretical Astrophysics, California Institute of
  Technology, Pasadena, California 91125, USA} 

\date{\today}

\begin{abstract}
  We discuss the properties of the effective-one-body (EOB) multipolar gravitational 
  waveform emitted by nonspinning black-hole binaries of masses $\mu$ and $M$ 
  in the extreme-mass-ratio limit $\mu/M=\nu\ll 1$.
  We focus on the transition from quasicircular inspiral to plunge, merger and ringdown.
  We compare the EOB waveform to a Regge-Wheeler-Zerilli  waveform  
  computed using the hyperboloidal layer method and extracted at null infinity.
  Because the EOB waveform keeps track analytically of most phase differences 
  in the early inspiral, we do not allow for any arbitrary time or phase shift between the waveforms.
  The dynamics of the particle, common to both wave-generation formalisms, is driven 
  by a leading-order ${\cal O}(\nu)$ analytically resummed radiation reaction. 
  The EOB and the Regge-Wheeler-Zerilli waveforms have an initial dephasing of about $5\times 10^{-4}$ rad and 
  maintain then a remarkably accurate phase coherence during the long inspiral 
  ($\sim 33$ orbits), accumulating only about $-2\times 10^{-3}$ rad until the last 
  stable orbit, \ie $\Delta\phi/\phi\sim -5.95\times 10^{-6}$. 
  We obtain such accuracy without calibrating the analytically-resummed EOB
   waveform to numerical data, which indicates the aptitude of the EOB waveform  for studies concerning the Laser Interferometer Space Antenna.
  We then improve the behavior of the EOB waveform around merger by 
  introducing and tuning next-to-quasicircular 
  corrections  in both the gravitational wave amplitude and phase. 
  For each multipole we tune only four next-to-quasicircular parameters by requiring 
  compatibility between EOB and Regge-Wheeler-Zerilli waveforms at the light ring. The resulting phase 
  difference around the merger time  is as small as $\pm 0.015$ rad, with a fractional 
  amplitude agreement of $2.5\%$. This suggest that next-to-quasicircular 
  corrections to the phase can be a useful ingredient in comparisons between EOB 
  and numerical-relativity waveforms.
\end{abstract}

 
\maketitle

\section{Introduction}

In the last few years numerical and analytical
relativity have demonstrated how to use 
information from the strong-field--fast-motion 
regime of coalescing black-hole binaries to build
accurate analytical models of their dynamics and 
of the gravitational radiation 
emitted~\cite{Buonanno:2007pf,Damour:2007yf,Damour:2007vq,Damour:2008te,Boyle:2008ge,Damour:2009kr,Damour:2009ic,Buonanno:2009qa,Damour:2008gu,Damour:2009sm,Pan:2009wj}.
Although numerical-relativity (NR) simulations of binary black holes 
have reached a high degree of accuracy and 
flexibility~\cite{Pretorius:2005gq,Gonzalez:2008bi,Pollney:2009yz,Lousto:2010tb,Lousto:2010ut}, 
a comprehensive spanning of the multidimensional parameter space 
remains prohibitive.
Analytical models are thus of fundamental importance to set up 
the bank of gravitational wave (GW) templates for detection.
The limiting case is given by extreme-mass-ratio inspirals (EMRIs) 
and mergers;  NR simulations simply can not access such regime and 
post-Newtonian (PN)  techniques are inaccurate at such velocities.
We need analytical models for the GW emission from EMRI
systems because they are primary target sources for the Laser Interferometer Space Antenna (LISA) 
and because their parameter space is very large~\cite{Gair:2004iv}.

The only analytical approach currently capable of accurately 
following the complete dynamics and providing waveforms 
(inspiral-plunge-merger-ringdown) of coalescing black-hole 
binaries is the effective-one-body (EOB) approach to the 
general relativistic two-body 
dynamics~\cite{Buonanno:1998gg,Buonanno:2000ef,Damour:2000we,Damour:2001tu,Buonanno:2005xu,Damour:2009ic}. 
The EOB formalism employs \emph{resummed} PN results 
(for dynamics and waveforms) in order to extend their validity 
in the strong-field--fast-motion regime, \ie in a region where they 
are inaccurate in their standard Taylor-expanded form.
In brief the analytical construction is based on (i) a dynamics governed by  
a resummed Hamiltonian and an expression for the mechanical angular 
momentum loss (the \emph{radiation reaction}) and (ii)
a waveform-generating algorithm which combines a prescription to resum 
the Taylor-expanded PN multipolar waveform up to the merger and 
a matching procedure to the quasinormal-mode (QNM) 
waveform to describe the postmerger phase (an oscillating black hole).

One key aspect of the EOB approach is its 
{\it flexibility}~\cite{Damour:2002vi}.
Although the formalism is based on analytical results known 
only at a given PN order, 
it is possible to take into account (yet uncalculated) higher-order
effects by means of suitable {\it flexibility parameters}.
These parameters may be determined (or just constrained)
by comparison with results from numerical-relativity simulations 
valid in the strong-field--fast-motion
regime. Several recent works~\cite{Damour:2007yf,Damour:2007vq,Damour:2008gu,
Damour:2009kr,Buonanno:2009qa,Pan:2009wj}
have shown how this tuning can be implemented to obtain analytical 
waveforms that match the numerical ones within numerical  errors. 
The tuned EOB formalism can then be used for parametric studies.

The Regge-Wheeler-Zerilli (RWZ) metric perturbation 
theory~\cite{Regge:1957td,Zerilli:1970se,Martel:2005ir,Nagar:2005ea}.    
is the natural tool to compute the GW emission from a system 
of two nonspinning black holes, of masses $\mu$ and $M$, 
in the extreme-mass-ratio limit (EMRL) $\nu\equiv \mu/M\ll 1$.
In this regime several numerical results can be used to calibrate the 
EOB dynamics and waveforms~\cite{Damour:1997ub,Damour:2007xr,Damour:2008gu,Yunes:2009ef,Yunes:2010zj}. 
In particular, recent gravitational--self-force
calculations~\cite{Barack:2009ey,Barack:2010tm} helped in putting
constraints on the functions entering the EOB conservative
dynamics~\cite{Damour:2009sm,Barack:2010ny}. 
The Regge-Wheeler-Zerilli perturbation theory has been used for many
years in the {\it Fourier domain} 
(see, for example, \cite{Berti:2010ce}
and references therein)
and {\it neglecting radiation-reaction effects}, 
since Davis, Ruffini, and Tiomno computed
the waveform emitted by a particle radially plunging into the black hole~\cite{Davis:1972ud}.
Only recently the RWZ approach has been extensively developed in
the {\it time domain}~\cite{Martel:2001yf,Martel:2003jj,Nagar:2004ns,Nagar:2005cj,
Sopuerta:2005gz,Canizares:2010ah} with the inclusion of the radiation-reaction 
force~\cite{Nagar:2006xv,Barack:2009ey,Barack:2010tm,Sundararajan:2010sr,Bernuzzi:2010ty}.

Time-domain simulations using the perturbation theory are efficient and accurate 
and complement NR simulations in EMRL. 
The first calculation of the complete gravitational waveform emitted during the
transition from inspiral to plunge, merger and ringdown in the EMRL was
performed in Ref.~\cite{Nagar:2006xv}, thanks to the combination of the
RWZ perturbation theory and the 2.5PN accurate (analytical) Pad\'e-resummed radiation-reaction 
force~\cite{Damour:1997ub}. Reference~\cite{Damour:2007xr} used 
that result as a target waveform to assess the performances of the corresponding 
EOB (resummed) analytical waveform. The comparison was restricted to the quadrupole 
case, $m=\ell=2$. The knowledge from that study was 
useful in subsequent EOB/NR waveform comparisons. The treatment of the analytical 
radiation reaction in the strong-field--fast-motion regime has been improved since 
then, thanks to a resummed and factorized form of the PN multipolar 
waveform~\cite{Damour:2007xr,Damour:2008gu,Fujita:2010xj}. 

In~\cite{Bernuzzi:2010ty} (hereafter paper I) two of us presented 
an accurate computation of the gravitational radiation generated by 
the coalescence of two circularized nonspinning black holes  
in the EMRL. The results were obtained with an improved version of the 
finite-difference code of~\cite{Nagar:2006xv,Damour:2007xr}, 
which implements the expression of the radiation-reaction force 
based on the (5PN-accurate) analytical waveform 
resummation of~\cite{Damour:2008gu}. 
The knowledge of the ``exact'' RWZ multipolar waveform opened 
the way to two main conclusions, extensively discussed in paper I: 
first, the computation of the final kick velocity imparted to the system 
by GW emission, $v^{\rm kick}/(c\nu^2)=0.0446$.
This value proved consistent with the corresponding one
extrapolated from a sample of numerical-relativity simulations~\cite{Gonzalez:2006md}
(see Fig.~7 and Tables~IV and V in paper I), as well as with the outcome 
of an independent calculation that relies on a different treatment of the 
radiation reaction~\cite{Sundararajan:2010sr}. 
Second, it was possible to show a very good agreement (at the $10^{-3}$ level)
between the mechanical angular momentum loss provided by the 
analytical expression of the radiation reaction and  
the GW angular momentum flux computed from the RWZ waveforms. 
This second result supports the consistency of our 
approach. Notably, the agreement between the two functions was excellent 
also {\it below} the last stable orbit (LSO) and almost along the entire plunge 
phase up to merger (see Figs.~8 and 9 of paper I).
The results of~\cite{Bernuzzi:2010ty} also turned out to be compatible 
with the first NR computation of binary black hole coalescence in the 
large-mass-ratio regime (1:100)~\cite{Lousto:2010ut}.
Recently~\cite{BNZ:2010} we further improved the 
RWZ approach of paper I by combining it with the hyperboloidal 
layer method~\cite{Zenginoglu:2010cq}. 
This approach brings two main benefits. 
First, it allows us to extract GWs at null infinity 
($\I^+$), thereby eliminating the gauge effects related to
the GW extraction at a finite radius.
In addition, because we evolve the RWZ equations on a 
smaller coordinate domain, we substantially improve 
the efficiency of our code. 

The aim of this paper is to perform a detailed comparison, multipole by
multipole, between the RWZ and the corresponding analytical waveforms
computed within the EOB approach.
For the particle dynamics both codes (RWZ and EOB) implement the 
resummed radiation-reaction force ${\cal F}_\varphi$ of~\cite{Damour:2008gu} 
updated to include  5PN-accurate terms also for subdominant multipoles. 
The latter come from the 5.5PN-accurate 
(Taylor-expanded) circularized multipolar waveform computed by 
Fujita and Iyer~\cite{Fujita:2010xj}. The particle dynamics is computed
within this 5PN-accurate (resummed) approximation and is  
the same in both codes. For simplicity, we decided not to improve 
it further by tuning the resummed flux entering the radiation 
reaction~\cite{Yunes:2009ef,Yunes:2010zj}.
For the waveform we compare the full multipolar structure 
up to $m=\ell=4$, going beyond the simple quadrupole contribution.

The waveform comparison brings new knowledge with respect to the flux 
comparison of paper I 
(see also~\cite{Damour:2007xr}) for two main reasons. 
First, we assess the performance of the resummed EOB waveform 
in describing the {\it phase} of each multipole. Second, we perform 
detailed analyses of the next-to-quasicircular (NQC) corrections 
that are needed in the  late-plunge phase. 
NQC effects are actually responsible for the 
differences in the EOB and RWZ fluxes in the strong-field--fast-motion regime, 
as it was pointed out in paper I (see Fig.~8 there and also the related
discussion in~\cite{Damour:2007xr}). 
At the waveform level, several
studies~\cite{Damour:2007xr,Damour:2009kr,Buonanno:2009qa,Pan:2009wj}
have demonstrated that NQC corrections to the EOB 
waveform (and radiation reaction) are needed to improve its agreement 
with the numerical one during the late-plunge and merger
phase. Previous works were restricted to
the quadrupole case. Two central new benefits of this paper are 
(i) the assessment of the complete multipolar EOB waveform in 
the EMRL during the transition from inspiral to plunge and merger; and (ii) 
the development of a robust procedure to tune NQC corrections 
to the gravitational wave amplitude and {\it phase}.

The paper is organized as follows. In Sec.~\ref{sec:RWZ} we summarize
the main features of our RWZ numerical target waveform described 
in detail elsewhere~\cite{Bernuzzi:2010ty,BNZ:2010}.
In Sec.~\ref{sec:EOBwaves} we describe the structure of the multipolar EOB 
waveform of~\cite{Damour:2008gu,Fujita:2010xj}, giving all the details of 
the implementation used here. 
In Sec.~\ref{sec:compare} we first present our results for the inspiral phase
and then describe the procedure to tune NQC parameters necessary to
improve the EOB waveform at merger. 
The discussion is based mainly on the $\ell=2$ multipoles.
In Sec.~\ref{sec:htotal} we assess the quality of the complete EOB multipolar 
waveform, discussing explicitly multipoles up to $\ell=4$.
We finally put together some concluding remarks in Sec.~\ref{sec:conclusions}.
Two appendixes are included to complement the information given in the main text.
Throughout this paper we use geometrized units with $c=G=1$.

\section{Regge-Wheeler-Zerilli waveforms}
\label{sec:RWZ}

We compute numerical waveforms at future null infinity via the time-domain 
RWZ perturbative approach introduced in~\cite{Nagar:2006xv} and 
improved in~\cite{Bernuzzi:2010ty,BNZ:2010}.
We perform a hyperboloidal evolution of the RWZ equations 
with a point-particle source modeling the smaller-mass black hole. 
The distributional $\delta$ function representing the particle is approximated
by a narrow Gaussian of finite width $\sigma\ll M$.
The dynamics of the particle is started using post-circular initial 
data as defined in~\cite{Buonanno:2000ef,Nagar:2006xv}, 
which generate negligible eccentricity at the beginning of the evolution.
The conservative part of the dynamics is governed by the 
$\nu\to 0$ limit of the EOB Hamiltonian (the Hamiltonian of a
particle on Schwarzschild spacetime) with the following, 
dimensionless variables: the relative separation 
$r=R/M$, the orbital phase $\varphi$, the orbital 
angular momentum $p_\varphi = P_\varphi/(\mu M)$, and the 
orbital linear momentum $p_{r_*}=P_{r_*}/\mu$, canonically 
conjugate to the tortoise radial coordinate separation $r_*=r + 2 \ln(r/2-1)$.
The expression for the analytical radiation-reaction force $\hat{\cal F}_\varphi$  
is described in~\cite{Damour:2008gu,Bernuzzi:2010ty} 
and has been updated with the new terms in the resummed waveform amplitude 
recently computed in~\cite{Fujita:2010xj} at fractional 5PN accuracy.
In the computation of $\hat{\cal F}_\varphi$ we sum over multipoles up to $\ell=8$
included. The dynamics is then computed by solving Eqs.~(1)-(7) of paper I.

Because of the different analytical approximation to the flux and
because of the hyperboloidal evolution, the RWZ waveforms 
employed here are {\it quantitatively new} with respect to those
of paper I. By contrast, at a qualitative level, there are no 
appreciable differences. In the following we 
shall comment only on the main new features of our perturbative 
approach. For a complete description of the method, the equations, and the notation, 
we refer to~\cite{Nagar:2006xv,Bernuzzi:2010ty,BNZ:2010}.

We adopt the hyperboloidal layer method for the RWZ equations 
to extract waves at future null infinity and to increase the
efficiency of our code~\cite{Zenginoglu:2007jw,Zenginoglu:2009ey,Zenginoglu:2010cq}.  
The essential ingredient of this approach is a suitable transformation 
of the standard Schwarzschild time coordinate $t$ in combination with
spatial compactification. We anticipate here part of the technical 
steps discussed in~\cite{BNZ:2010}. 

We solve the RWZ equations, written in a general coordinate system~\cite{Sarbach:2001qq},
using coordinates $(\tau,\rho)\in \mathbb{R}^+\times \left[R_*^-,S\right]_{R_*^+}$.
The coordinates $(\tau,\rho)$ coincide with $(t,r_*)$ in a domain 
$D^{-+}_{r_*}=[R_*^-,R_*^+]$, that entirely includes the motion of the particle. 
The RWZ equations on $D^{-+}_{r_*}$ have then the same form as in paper I. 
The spatial compactification sets in at the interface $\rho = r_* = R_*^+$ in a  
sufficiently differentiable way. The compactifying coordinate $\rho(r_*)$ maps the
infinite $r_*$ domain $D^+_{r_*}=[R_*^+,\infty)$ to the finite $\rho$ domain 
$D^+_\rho=[R_*^+,S]$, where $S>R_*^+$ is a constant.
A new time coordinate is introduced according to the prescription 
that the timelike Killing field is left invariant, $\partial_t =
\partial_\tau$, which implies 
\be
\label{eq:tau_transf}
\tau = t-h(r_*) \ .  
\ee
The height function $h(r_*)$ is related to $\rho(r_*)$ by the 
condition that the representation of outgoing null rays is 
left invariant
\be
\label{eq:tau_condition}
t - r_*= \tau - \rho \ .
\ee
Equations~\eqref{eq:tau_transf} and ~\eqref{eq:tau_condition} imply 
$h(r_*)=r_*-\rho(r_*)$ that, together with the choice of a 
sufficiently smooth spatial compactification, determines a (future)
hyperboloidal foliation. Thus the surface $\rho=S$ corresponds to
future null infinity $\I^+$, and outgoing waves are 
evenly resolved in the compactifying coordinate $\rho$. 
Our numerical domain reads $[R_*^-,S]_{R_*^+}=[-50,70]$ 
with the interface at $R_*^+=50$ and is covered by 
$N=3001$ grid points~\cite{BNZ:2010}. 

\begin{figure}[t]
\begin{center}
\includegraphics[width=0.45\textwidth]{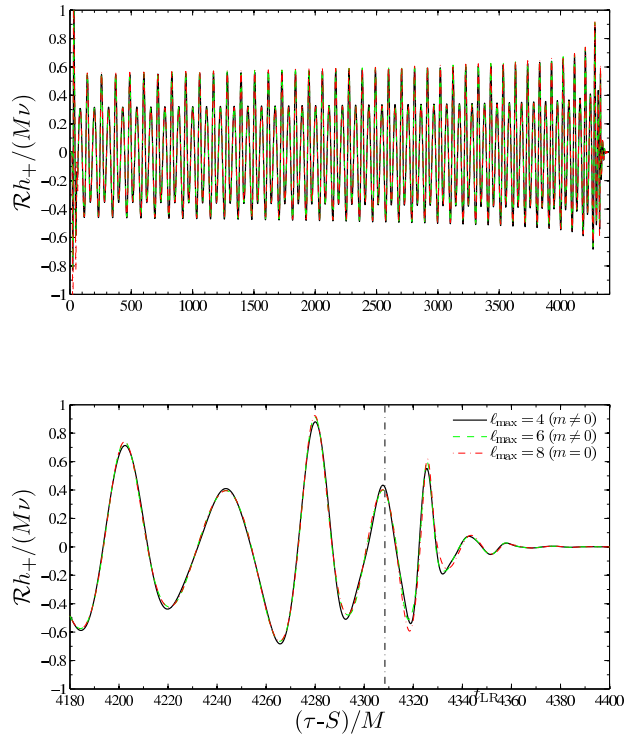}
\caption{\label{fig:RWZ_exact} Multipolar ``convergence'' of the $\R h_+/(M\nu)$ polarization 
 of the Regge-Wheeler-Zerilli multipolar waveform. Top panel: The complete wave train 
 ($\sim 37$ cycles). Bottom panel: Impact of subdominant multipoles around the merger time.
 The vertical dashed line indicates the light-ring crossing.}
\end{center} 
\end{figure}

The RWZ-based approach we use to compute the numerical 
target waveforms relies on certain approximations. 
Our first approximation is to consider the dissipative radiation reaction at
leading order in the mass ratio $\nu$,
neglecting higher-order
corrections that enter both the conservative and the nonconservative
parts of the dynamics~\cite{Barack:2009ey,Barack:2010tm,Blanchet:2009sd}. 
If one is interested in computing very long inspiral waveforms, 
these higher-order effects
must be properly taken into account for LISA-related data analysis 
(for example, using the EOB formalism~\cite{Yunes:2009ef,Yunes:2010zj,Bernuzzi:2010ty}).  
The computation of long inspiral waveforms with the ${\cal O}(\nu)$ radiation reaction 
is affected by systematic uncertainties that depend on the mass ratio $\nu$ 
and on the integration time. 
By contrast, when one focusses only on the late-time part of the 
waveform, \ie the one corresponding to the ``quasigeodesic'' plunge, 
merger, and ringdown, one can extrapolate finite-$\nu$ results to the $\nu= 0$ limit.
In paper I we followed this procedure for dynamical quantities like the 
kick velocity or the energy emitted during the plunge, and we showed 
that the impact of dealing with a finite value of $\nu$ becomes less and 
less important when $\nu\leq 10^{-3}$. 
Our second approximation is that the expression
for the radiation-reaction force $\hat{{\cal F}}_\varphi$ is based on resummed PN results 
for circularized binaries and, as such, it neglects 
nonquasicircular corrections that explicitly depend on the 
radial momentum and its derivatives. The accuracy of this second 
approximation can be checked only \emph{a posteriori}, but it is typically 
quite good also during the plunge~\cite{Damour:2007xr,Bernuzzi:2010ty}.

On the basis of the considerations of paper I, we choose the mass ratio 
$\nu=10^{-3}$ and the initial separation $r_0=7M$ for the present EOB and RWZ
comparison. These values guarantee a long inspiral  ($\sim 37$ orbits) 
as well as an accurate representation of
the late-time waveform (see its mild 
dependence on $\nu$ when moving from $\nu=10^{-3}$ 
and $10^{-4}$ in Fig.~4 of paper I). We follow here the same normalization 
convention for the RWZ $\Psi_{\ell m}^{(\rm e/o)}$ master functions 
[with (e)ven-parity and (o)dd-parity modes] as in paper I, \ie
\be
\label{eq:hplus_cross}
\R(h_+ -\ii h_\times) = \sum_{\ell=2}^{\ell_{\max}}\sum_{m} \sqrt{\dfrac{(\ell+2)!}{(\ell-2)!}} \left(\Psi_\lm^{(\rm e)} + \ii \Psi_\lm^{(\rm o)}\right){}_{-2}Y_{\ell m},
\ee
where $\R$ is the distance from the source, $\ell_{\max}$ is the maximum number
of multipoles that we consider, and ${}_{-2}Y_{\ell m}\equiv{}_{-2}Y_{\ell m}(\Theta,\Phi)$ 
are the $s=-2$ spin-weighted spherical harmonics computed 
in the convention of Ref.~\cite{Kidder:2007rt}.
Figure~\ref{fig:RWZ_exact} displays various multipolar approximations to the  
$\R h_+/(M\nu)$ polarization of the RWZ waveform
along the fiducial direction $(\Theta,\Phi)=(\pi/4,0)$.
The waveforms are shown versus retarded time at $\I^+$, $\tau-S$.
The dashed line in the figure (red online) refers to the complete waveform 
obtained summing the multipoles up to $\ell=8$. 
The other two lines depict the partial contributions to the total waveform 
up to $\ell=4$ (solid line) or $\ell=6$ (dash-dotted line) and
neglecting the $m=0$ multipoles. The bottom panel of the figure is a close-up
on the waveform around the conventional ``merger'' time, 
\ie the time $t_{\rm LR}$ at which the particle crosses 
the light-ring $r=3M$ (vertical dash-dotted line).
The figure gives us a visual idea of the impact of the various multipoles on
the accuracy of the total waveform, and it indicates that 
$\ell_{\max}=4$ (with only $m\neq 0$) gives us a good approximation, 
especially up to $t_{\rm LR}$. For this reason, and to keep the discussion 
sufficiently simple, we shall fix $\ell_{\max}=4$ and consider 
only $m\neq 0$ multipoles to compute the RWZ target waveform to be compared 
with the EOB analytical waveform. We will discuss the fine details of the complete
 RWZ waveform in~\cite{BNZ:2010} (see also~\cite{Bernuzzi:2010ty}).

\section{Effective-one-body resummed multipolar waveform}
\label{sec:EOBwaves}

In this section we review the structure of the EOB 
waveform~\cite{Damour:2008gu,Damour:2009kr,Damour:2009ic,Fujita:2010xj}.
The EOB-resummed multipolar waveform can be split into two parts: 
the inspiral-plus-plunge-and-merger ({\it insplumerg}) waveform computed
during the dynamics of the particle up to merger
and the {\it ringdown} waveform that describes the post merger 
waveform as a superposition of black-hole QNMs.
A simplified and efficient representation of the transition
between the late-plunge and the ringdown regimes 
is accomplished by {\it matching} the insplumerg waveform to the 
ringdown waveform\footnote{We introduced the 
nomenclature ``insplumerg waveform'' to indicate the part of the 
EOB waveform that is usually called  {\it insplunge waveform} in 
the literature~\cite{Damour:2009kr,Damour:2009ic}. The reason for 
this choice is twofold: first, because it includes NQC corrections
that become relevant essentially only around the merger time, and  
second, because it needs compatibility conditions with the RWZ 
waveform {\it around merger} to be fully determined.}.
The complete EOB multipolar waveform reads
\begin{align}
\label{eq:eobwvf}
h_\lm^{\rm EOB}\, {}^{(\epsilon)}(t;a_i^\lm,\sigma_{n\ell}^{\pm}) &=
\theta(t_m - t)\, h_\lm^{\rm insplumerg}\, {}^{(\epsilon)}(t;a_i^\lm) \nonumber\\
&+ \theta(t-t_m)\, h_\lm^{\rm ringdown}\,{}^{(\epsilon)}(t;\sigma_{n\ell}^\pm)\ ,
\end{align}
where $\epsilon$ denotes the parity, \ie even ($\epsilon=0$) for 
mass generated multipoles and odd ($\epsilon=1$) for current generated 
ones\footnote{For notational consistency with previous analytical 
work~\cite{Damour:2008gu} we label even- and odd-parity modes 
with $\epsilon=0$ and $\epsilon=1$, respectively, while 
in Eq.~\eqref{eq:hplus_cross} above  we used the labelling (e) and (o).}.
Since the particle motion is planar, $\epsilon$ is equal to the parity of the sum~$(\ell+m)$.
Furthermore, we explicitly highlight the dependence on the NQC parameters $a_i^{\ell m}$ 
and the QNM (complex) frequencies $\sigma_{n\ell}^{\pm}$. 

The ringdown waveform is written as
\be
h_{\lm}^{\rm ringdown}\, {}^{(\epsilon)}=\sum_n C_{n\ell m}^{+}e^{-\sigma^+_{n\ell} t} 
+ \sum_n C_{n\ell m}^{-}e^{-\sigma^-_{n\ell}t},
\ee
where, following Ref.~\cite{Damour:2007xr}, we use the notation 
$\sigma_{n\ell}^{\pm}=\alpha_{n \ell}\pm \ii \omega_{n\ell}$ 
for the positive and negative QNM frequencies and $C_{n\ell m}^{\pm}$ for the 
corresponding amplitudes (note that, for simplicity, we omitted here 
the parity index $\epsilon$).
Here $\omega_{n\ell}$ and $\alpha_{n\ell}$ indicate the frequency and the inverse damping 
time of each mode respectively, and $n=0,1,2,\dots N-1$ label the overtone 
number ($n=0$ denoting the fundamental mode).

The insplumerg waveform can be written as the product of several factors.
We factorize the NQC correction as
\be
\label{eq:insplumerg}
 h_\lm^{\rm insplumerg}\, {}^{(\epsilon)}(t;a_i^\lm) = h_\lm^{\rm insplunge} {}^{(\epsilon)} h_\lm^{\rm NQC}(a_i),
\ee
and the {\it insplunge waveform} is given as the product of the Newtonian contribution
and a PN (resummed) correction by
\be
\label{eq:hlm}
h_\lm^{\rm insplunge}\, {}^{(\epsilon)} \equiv 
h_\lm^{(N,\epsilon)}(x)\hat{h}^{(\epsilon)}_{\ell m}. 
\ee
The Newtonian contribution is given by 
\be
h_{\ell m}^{(N,\epsilon)}=
\dfrac{M\nu}{\cal R}n_{\ell m}^{(\epsilon)}c_{\ell + \epsilon}(\nu)x^{(\ell
  +\epsilon)/2} Y^{\ell -\epsilon,-m}\left(\dfrac{\pi}{2},\varphi \right),
\ee 
where $Y^{\ell m}(\theta,\phi)$ are the usual scalar spherical harmonics 
(computed on the equatorial plane $\theta=\pi/2$), 
and the numerical coefficients $n_{\ell m}^{(\epsilon)}$ and $c_{\ell +\epsilon}(\nu)$ 
are explicitly given by Eqs.~(5)-(7) of Ref.~\cite{Damour:2008gu}.
Because we work here in the EMRL ($\nu\to 0$), we pose $c_{\ell +\epsilon}(0)$,
leaving only the overall factor $\nu$. Following Ref.~\cite{Damour:2009kr} 
(consistently with Ref.~\cite{Damour:2007xr,Damour:2006tr}), the argument $x$ 
in the Newtonian prefactor of Eq.~\eqref{eq:hlm} is taken as
\be
x=v_\varphi^2=(r\Omega)^2 \ ,
\ee 
where $\Omega$ is the orbital frequency.
This choice is preferable to $ x= x_{\rm circ}\equiv\Omega^{2/3}_{\rm circ}$ 
due to the violation of the circular Kepler's law during the plunge phase.
The quantity 
\be
\label{eq:hresum}
\hat{h}^{(\epsilon)}_{\ell m}\equiv \hat{S}_{\rm eff}^{(\epsilon)} T_{\ell m} e^{\ii \delta_{\ell m}}
(\rho_{\ell m})^{\ell }
\ee
represents a factorized (and resummed) version of all 
the PN corrections to the waveform.
It is given as the product of four factors: the $\mu$-normalized effective
source $\hat{S}_{\rm eff}^{(\epsilon)}$, the tail factor $T^{\ell m}$ 
that resums an infinite number of leading logarithms entering the 
{\it tail effects}, the supplementary phase $\delta_{\ell m}$,
and the residual modulus correction $\rho_{\ell m}$.
The even-parity effective source $\hat{S}_{\rm eff}^{(0)}$ is given by
the $\mu$-normalized Hamiltonian of the system computed along the dynamics, 
while the odd-parity one $\hat{S}_{\rm eff}^{(1)}$ is given by the corresponding 
(Newton-normalized) angular momentum.
The explicit expression of the tail factor as a function of
the orbital frequency $\Omega$ reads 
\be
\label{eq:tail}
T_{\ell m}(\Omega) = \dfrac{\Gamma(\ell+1-2\ii \hk)}{\Gamma(\ell +1)}
e^{\pi\hk} e^{2\ii \hk \ln(2 k r_{0s})},
\ee
where $k=m\Omega$, $\hk=Mk$. Following~\cite{Fujita:2010xj}, we denote
by $r_{0s}$ the quantity that was previously denoted as $r_0$ in~\cite{Damour:2008gu},
and we choose $ r_{0s} = r_0 =  2M/\sqrt{e}$ 
for consistency between the computations of the phase of~\cite{Fujita:2010xj} 
and~\cite{Damour:2008gu}. This value is chosen to match PN
results (in harmonic coordinates) with black-hole perturbation results in
Schwarzschild coordinates used here. 

Finally, the factor $h_{\ell m}^{\rm NQC}(a_i^{\ell m})$ is 
a correction that models the noncircular
effects in the waveform. 
This effective term allows us to analytically compute the waveform beyond the circular approximation. 
The necessity of using a NQC corrective factor 
to the waveform {\it amplitude} (usually denoted as $f^{\rm NQC}$) was first 
pointed out in~\cite{Damour:2007xr} and then used (with variations) in 
several studies~\cite{Damour:2007vq,Damour:2008te,Buonanno:2009qa,Damour:2009kr,Pan:2009wj}.
The inclusion of such a NQC correction proved necessary to improve the closeness
of the $m=\ell=2$ EOB waveform to the numerical one during the late-plunge and
merger dynamics. 
Building on this knowledge we extend the use  
of $h_{22}^{\rm NQC}$~\cite{Damour:2007xr,Damour:2009kr} 
to the other multipoles, with the important difference that we consider also
corrections to the waveform {\it phase}~\cite{Damour:2007xr,Buonanno:2009qa}.
This generalized NQC factor takes the form
\begin{align}
\label{eq:fNQC}
h^{\rm NQC}_{\ell m}(a_i^{\ell m}) 
= \left(1 + a_1^{\ell m}\dfrac{p_{r_*}^2}{(r\Omega)^2}+a_2^{\ell m}\dfrac{\ddot{r}}{r\Omega^2}\right)\nonumber \\
\times \exp\left[\ii \left(a_3^{\ell m} \dfrac{p_{r_*}}{r\Omega} + a_4^{\ell m} \dfrac{\dot{\Omega}}{\Omega^2}\right)  \right],
\end{align}
where the  $a_i^\lm$'s are NQC flexibility parameters that have to be determined, multipole by multipole, 
by imposing some compatibility conditions between the  EOB and RWZ waveforms around the merger time.

\section{Comparing EOB and RWZ waveforms}
\label{sec:compare}
In this section we present the comparison between RWZ and EOB
waveforms. 
In the first part we assess the performances of the resummed
{\it insplunge waveform}, Eq.~\eqref{eq:hlm}, to describe the 
inspiral up to the LSO crossing. The insplunge waveform 
includes neither NQC corrections nor matching to QNMs. 
Nonetheless it shows a very good agreement with RWZ during 
the whole inspiral up to (and even below) the LSO.
In the second part we focus on the transition from inspiral to plunge,
merger and ringdown, and we compare to the RWZ waveform the three different 
analytical representations of the resummed multipolar waveform introduced above.
Specifically we consider the same insplunge waveform mentioned above; 
the {\it insplumerg waveform} that includes only the 
NQC corrections (to both phase and amplitude), 
Eq.~\eqref{eq:insplumerg}; and the full EOB 
waveform, Eq.~\eqref{eq:eobwvf}, with NQC corrections 
and with QNMs matching. We emphasize the necessity of introducing 
NQC corrections to the resummed EOB waveform and we propose compatibility
conditions with the RWZ waveform to determine the NQC parameters
$a_i^\lm$. The procedure we discuss is equally
robust for all multipoles and does not require any relative time and
phase shifting of the waveforms, or any hand-tuning of parameters. 

We present our comparison consistently with the notation of paper I; \ie we
use ``Zerilli-normalized'' metric multipoles $\Psi_{\ell m}$. These are related
to the $ \R h_{\ell m}$ metric multipoles as
\begin{equation}
\Psi^{\rm X,(\epsilon)}_{\ell m} = \ii^\epsilon N_\ell \R h^{\rm X(\epsilon)}_{\ell m},
\end{equation}
where $N_\ell =1/\sqrt{(\ell +2)(\ell +1)\ell (\ell -1)}$ and the 
label ${\rm X}$ stays for either insplunge, insplumerg, or EOB.
Note that this convention implies a phase shift of $\pi/2$ between
the odd-parity multipoles $\R h_{\ell m}^{(1)}$ and $\Psi_\lm^{(1)}$.
The EOB and RWZ multipolar waveforms are computed from the same
dynamics; the RWZ one is typically extracted at  $\I^+$ and shown 
versus the corresponding retarded time $\tau-S$. 
The EOB waveforms is parametrized by the dynamical time $t$ 
that will always be used as the reference time axis. 
Finally, the phases $\phi_\lm$ and the amplitudes $A_\lm$ of 
the (RWZ or EOB) complex numbers  $\Psi_\lm$ are defined with the convention  
$\Psi_\lm=A_\lm e^{-\ii\phi_{\ell m}}$.

\begin{figure}[t]
\begin{center}
\includegraphics[width=0.45\textwidth]{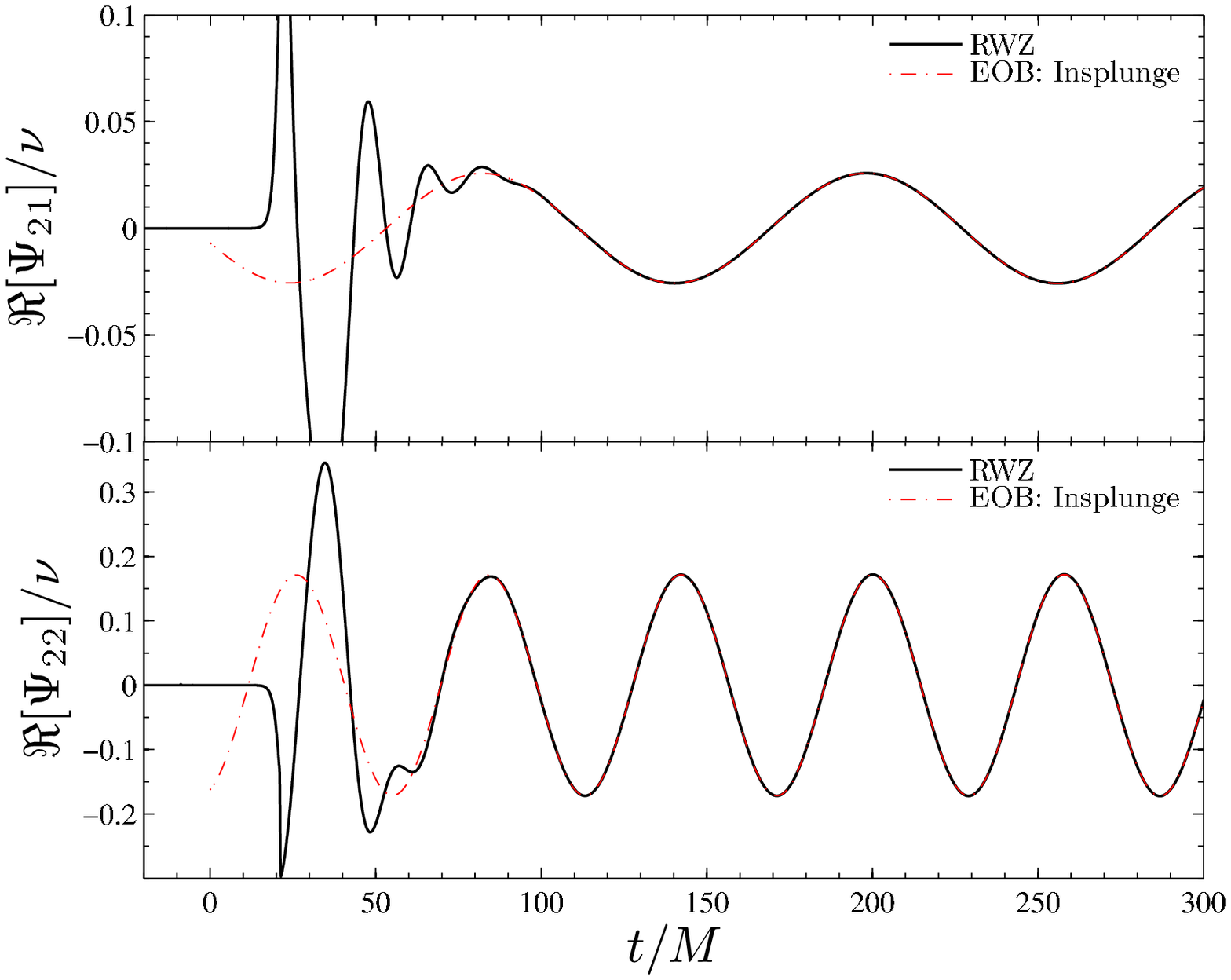}\\
\vspace{5 mm}
\includegraphics[width=0.45\textwidth]{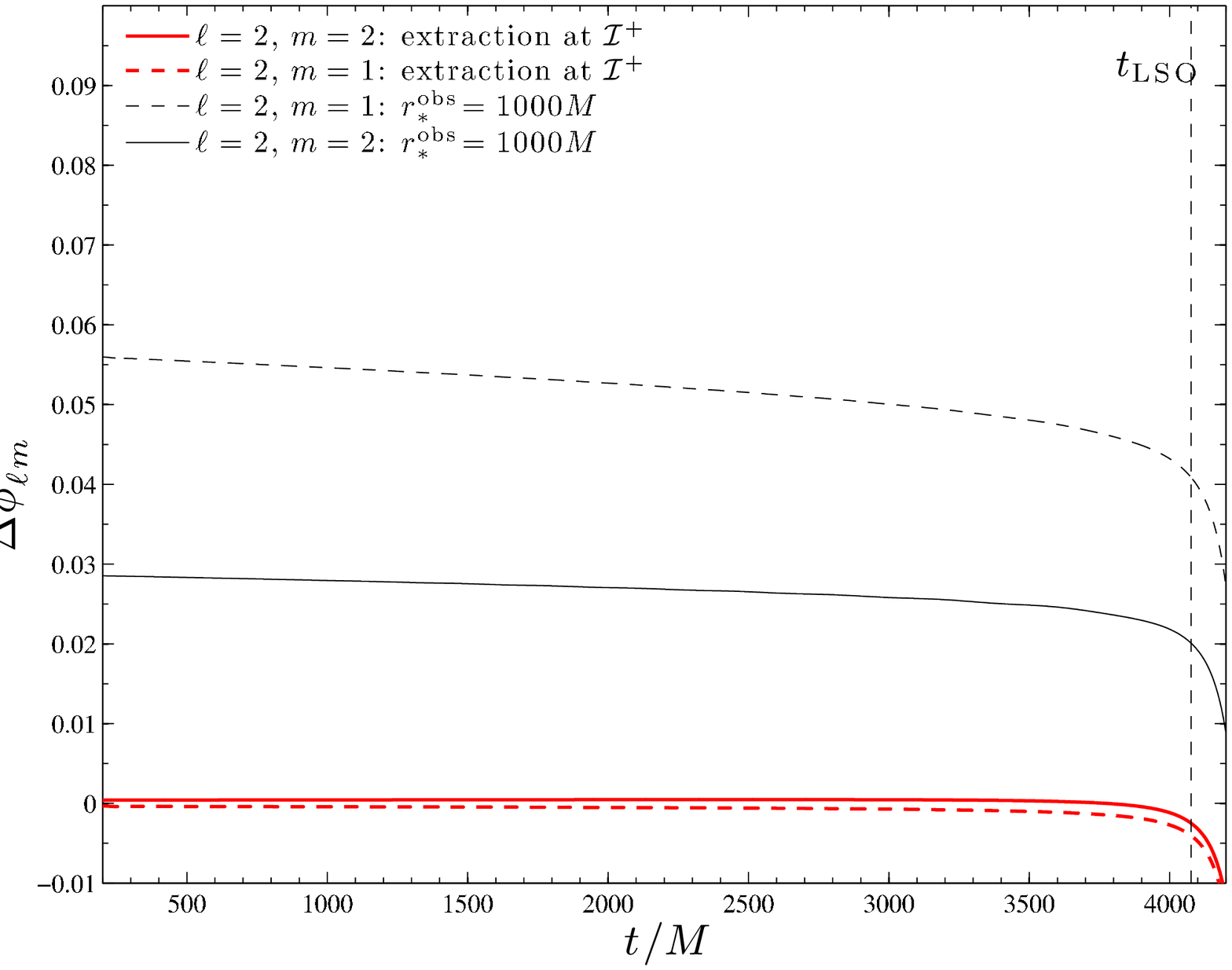}
\caption{\label{fig:early_time}Testing the waveform resummation
  for $\ell=2$ at the beginning of the inspiral.
  Top panel: Insplunge and RWZ waveforms extracted at $\I^+$.
  Bottom panel: The phase differences 
  $\Delta\phi^{\rm EOBRWZ}_{\ell m}=\phi^{\rm EOB}_{\ell m}-\phi^{\rm RWZ}_{\ell m}$,
  for RWZ waveforms measured at $\I^+$, are contrasted with the
  corresponding ones for RWZ waveforms measured at a finite extraction 
  radius $r_*^{\rm obs}=1000M$.}
\end{center} 
\end{figure}

\begin{table}[t]
  \caption{\label{tab:LSO_insplunge}Phase difference $\Delta\phi^{\rm insplunge}\equiv\phi^{\rm EOB}_{\ell m}-\phi^{\rm RWZ}_{\ell m}$ 
   and relative amplitude difference $(\Delta A/A)^{\rm insplunge} = (A_\lm^{\rm EOB}- A_\lm^{\rm RWZ})/A_\lm^{\rm RWZ}$ 
   between insplunge EOB and RWZ waveforms computed at $t_0=500M$ and at the LSO crossing, $t_{\rm LSO}=4076.1M$, 
   for all the ($m\neq 0$) multipoles with $2\leq \ell\leq 4$.}  
  \begin{center}
    \begin{ruledtabular}
    \begin{tabular}{llcccc}
      $\ell$ & $m$ & $\Delta\phi_{500M}^{\rm insplunge}$ & $\Delta\phi_{\rm LSO}^{\rm insplunge}$ & $(\Delta A/A)^{\rm insplunge}_{500M}$ & $(\Delta A/A)^{\rm insplunge}_{\rm LSO}$ \\
      \hline 
      2& 1&	 -3.91 $\times 10^{-4}$ & 	 -4.04$\times 10^{-3}$&	 -1.16$\times 10^{-4}$&	 -9.62$\times 10^{-4}$ \\
      2& 2&	  4.16 $\times 10^{-4}$ &	 -2.48$\times 10^{-3}$&	 -1.40$\times 10^{-3}$&	 -2.72$\times 10^{-3}$\\
      \hline 
      3& 1&	 -1.74 $\times 10^{-3}$ &	 -1.86$\times 10^{-2}$&	 -5.56$\times 10^{-4}$&	 -4.6$\times 10^{-3}$\\
      3& 2&	 -3.29 $\times 10^{-4}$ & 	 -4.65$\times 10^{-3}$&	 7.54$\times 10^{-4}$&	 9.54$\times 10^{-4}$\\
      3& 3&	 3.19  $\times 10^{-4}$ &	 -3.76$\times 10^{-3}$&	 -1.98$\times 10^{-3}$&	 -4.10$\times 10^{-3}$\\
      \hline 
      4& 1&	 -2.06 $\times 10^{-3}$ &	 -2.36$\times 10^{-2}$&	 -2.96$\times 10^{-4}$&	 -8.02$\times 10^{-3}$\\
      4& 2&	 -1.35 $\times 10^{-3}$ & 	 -1.51$\times 10^{-2}$&   -9.55$\times 10^{-4}$&	 -3.98$\times 10^{-3}$\\
      4& 3&	 2.78  $\times 10^{-3}$ & 	 5.69$\times 10^{-4}$&	 1.15$\times 10^{-3}$&	 1.80$\times 10^{-3}$\\
      4& 4&	 4.22  $\times 10^{-4}$ &	 -4.40$\times 10^{-3}$&	 3.42$\times 10^{-3}$&	 7.1$\times 10^{-3}$\\
    \end{tabular}
  \end{ruledtabular}
\end{center}
\end{table}

\subsection{Quasiadiabatic inspiral}

Let us focus first on the quality of the resummed 
waveform during the long ($\sim 37$ orbits) quasiadiabatic inspiral. 
We identify here the end of the inspiral 
(and the beginning of the plunge) as the time $t_{\rm LSO}=4076.2M$ at 
which the particle crosses the last stable orbit $r_{\rm LSO}=6M$.
This is clearly a convention because the transition from the
inspiral to plunge is a  blurred process that occurs 
around $r_{\rm  LSO}$~\cite{Buonanno:1998gg,Buonanno:2000ef,Ori:2000zn}. 

The upper panel of Fig.~\ref{fig:early_time} displays the early-time
evolution of the real part of the $\ell=2$ EOB multipoles (dashed lines) 
(top $m=1$; bottom $m= 2$) together with the RWZ ones (solid lines). 
After the initial unphysical transient the plot shows a remarkably good
agreement between phases and amplitudes of the two waveforms.
The time evolution of the corresponding phase differences
(during the complete inspiral)
$\Delta\phi^{\rm EOBRWZ}_{\ell m}(t)=\phi^{\rm EOB}_{\ell m}(t)-\phi^{\rm RWZ}_{\ell m}(t)$
is shown (as thicker lines, red online) in the lower panel of Fig.~\ref{fig:early_time}.
The vertical dashed line in the lower panel of the figure (with label $t_{\rm LSO}$)
indicates the LSO-crossing time. 
At the beginning, say for $t/M < 500$ (first 4 orbital cycles), the magnitude of 
both $\Delta\phi^{\rm EOBRWZ}_{22}(t)$ and $\Delta\phi^{\rm EOBRWZ}_{21}(t)$
is below $10^{-3}$. More precisely, at $t=t_0=500M$, we have  $\Delta\phi^{\rm EOBRWZ}_{22}(t_0)=4.16\times 10^{-4}$
and  $\Delta\phi^{\rm EOBRWZ}_{21}(t_0)=-3.98\times 10^{-4}$.
Such a small dephasing is compatible with the expected uncertainty 
on the 4.5PN-accurate phases $\delta_{\ell m}$ in Eq.~\eqref{eq:hresum}. 
By evaluating the last (the 4.5PN ones) terms in Eqs.~(5.8a)-(5.8b) 
of~\cite{Fujita:2010xj} at the initial position $r_0=7M$,
we estimate $\delta_{22}^{\rm 4.5PN}\sim4.16\times 10^{-3}$ and
$\delta_{21}^{\rm 4.5PN}\sim5.27\times 10^{-4}$.
Even if the phase difference grows by an order of magnitude on
the interval $[t_0,t_{\rm LSO}]$, it remains small at $t_{\rm LSO}$.
Table~\ref{tab:LSO_insplunge} accounts for the complete information,
multipole by multipole, about phase and fractional amplitude differences,
$(\Delta A/A)_\lm = (A_\lm^{\rm EOB}- A_\lm^{\rm RWZ})/A_\lm^{\rm RWZ}$, both
at $t_0$ and at $t_{\rm LSO}$.

When we sum together all the multipoles in
the complete waveform, Eq.~\eqref{eq:hplus_cross}, we initially obtain 
$\Delta\phi^{\rm insplunge}_{500M}=5\times 10^{-4}$ (corresponding to total 
GW phase $\phi=51.74$ rad), that becomes as large as $-2\times 10^{-3}$ at 
the LSO ($\phi=472.1$ rad). 
This means that the complete EOB insplunge waveform has dephased from the 
numerical one by only about $-2.5\times 10^{-3}$  rad over the $\sim 420$ rad 
($\sim 33$ orbits) of GW phase evolution on the interval $[t_0,t_{\rm LSO}]$,
which yields $\Delta\phi^{\rm EOBRWZ}/\phi^{\rm RWZ}=-5.95\times 10^{-6}$.

We point to one of the main findings of Ref.~\cite{BNZ:2010} in the lower panel of Fig.~\ref{fig:early_time}. The figure depicts the phase differences 
to numerical waveforms extracted at the {\it finite radius}\footnote{We checked
this result also using the
numerical setup of paper I, \ie with evolution along Cauchy time surfaces 
on a finite-size $r_*$ domain $r_*\in[-500,1500]$ and artificial 
boundaries (with Sommerfeld's outgoing boundary conditions) instead 
of a hyperboloidal layer. A thorough discussion of these effects will
be given in Ref.~\cite{BNZ:2010}.} $r_*^{\rm obs}=1000M$ and displayed
versus the corresponding observer retarded time.
The plot clearly shows that the values of $\Delta\phi_{\lm}^{\rm EOBRWZ}$
obtained with waves extracted at a finite radius (though very large)
are approximately 2 orders of magnitude larger than those obtained
with waves extracted at null infinity. The phenomenon discussed here 
in the $\ell=2$ case remains the same for the other multipoles. 
In particular, the largest and the smallest dephasing is for the $m=1$ and 
the $m=\ell$ multipoles respectively.
The dephasing might grow (up to the $\sim 0.1$ rad level) for 
$\ell>2$. The small dephasing of the waveform from the hyperboloidal layer 
calculation is the result of accurate wave extraction at null 
infinity~\cite{BNZ:2010}. At the present stage, this simple plot teaches 
us a very useful lesson: Extracting waves at $1000M$, even in the EMRL, 
may introduce dephasings that are considerably {\it larger} than the 
ones expected on the basis of the analytical knowledge of the GW phase. 
\emph{A priori} one expects such a phenomenon to be equally relevant, or even 
more important, in the comparable mass-ratio case due to the smaller 
PN accuracy at which the $\delta_{\ell m}$'s are known 
when $\nu\neq 0$~\cite{Damour:2008gu}. 

\begin{figure*}[t]
\begin{center}
  \includegraphics[width=0.45\textwidth]{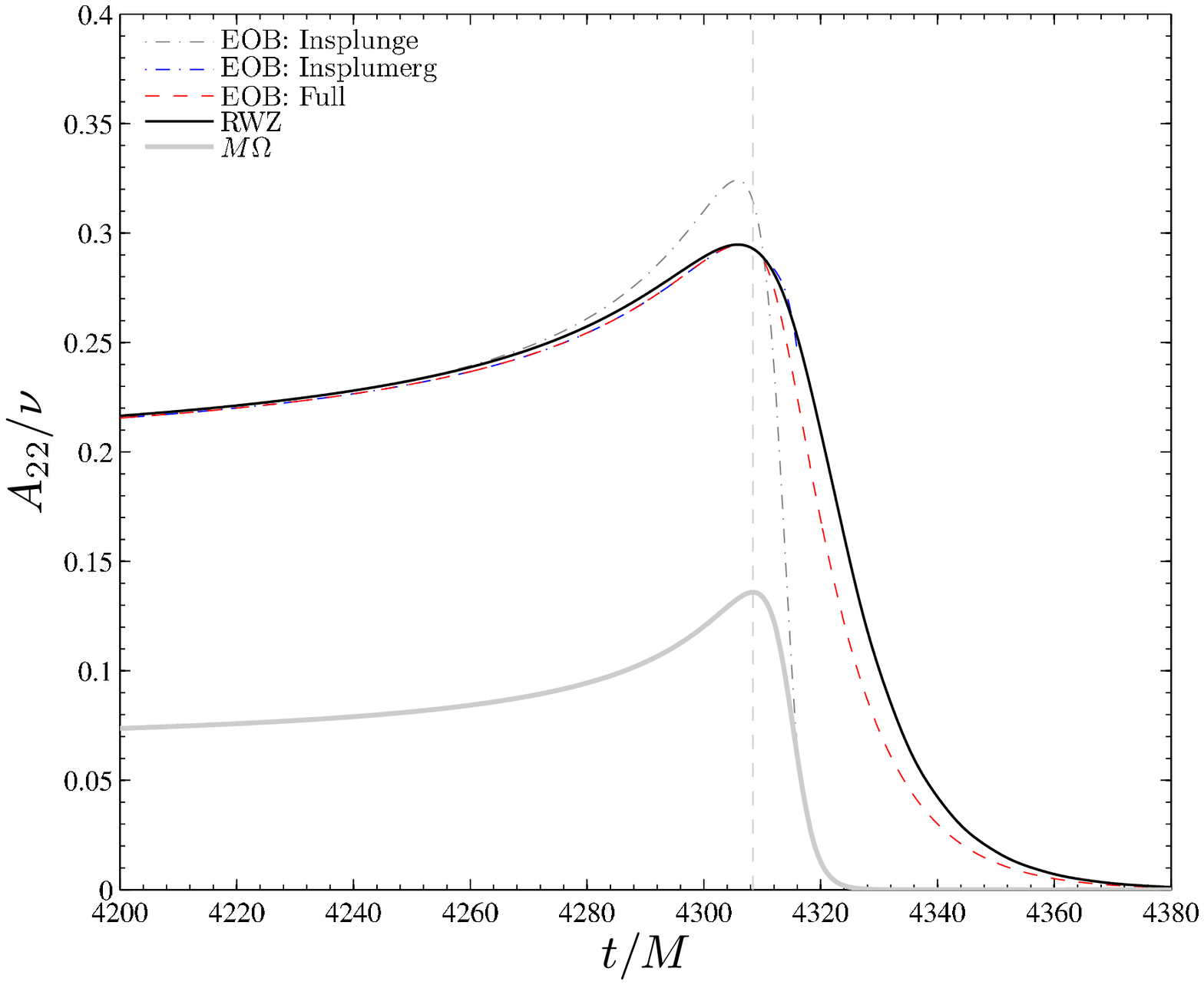}
\hspace{5 mm}
  \includegraphics[width=0.45\textwidth]{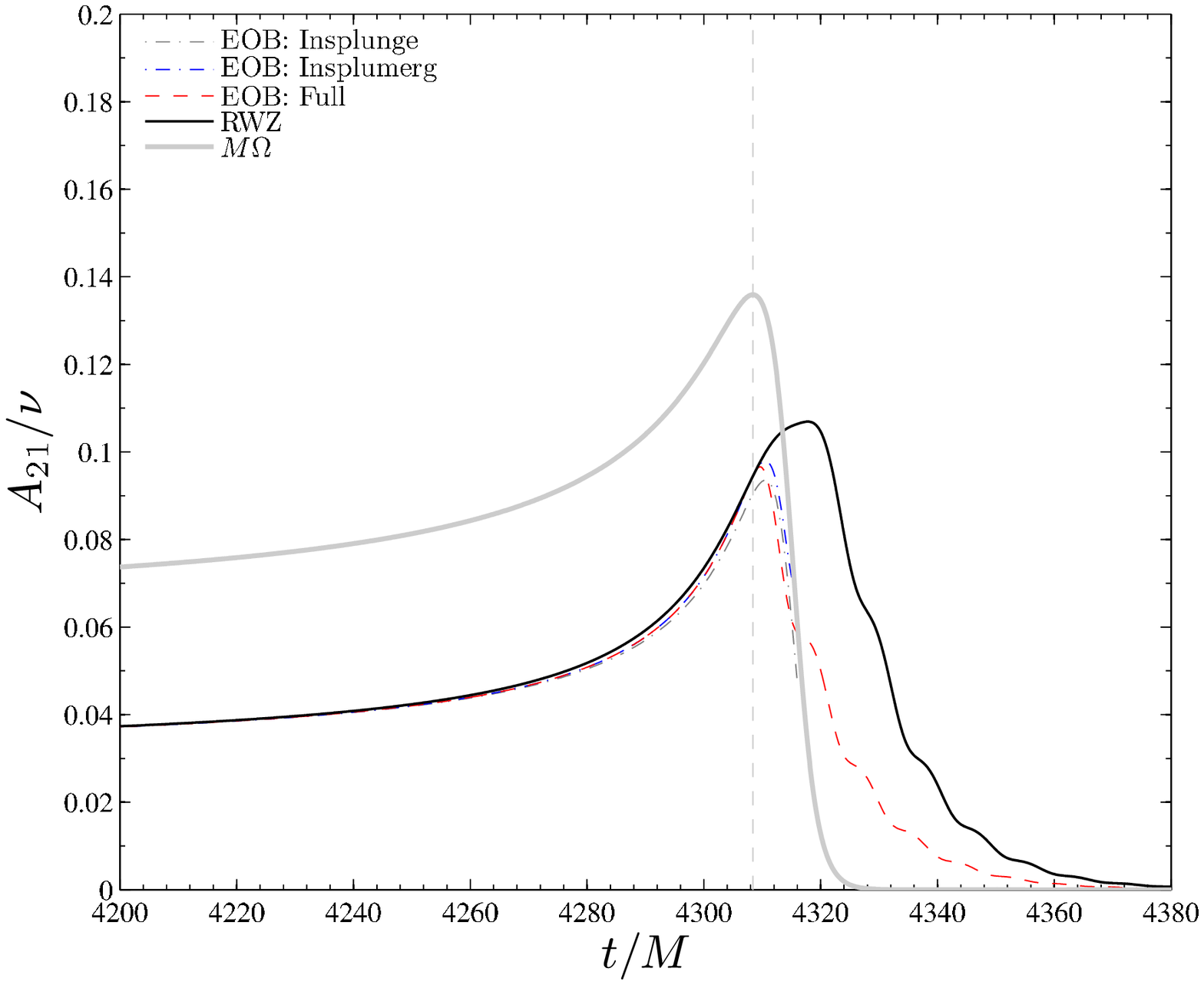}\\
\vspace{5 mm}
  \includegraphics[width=0.45\textwidth]{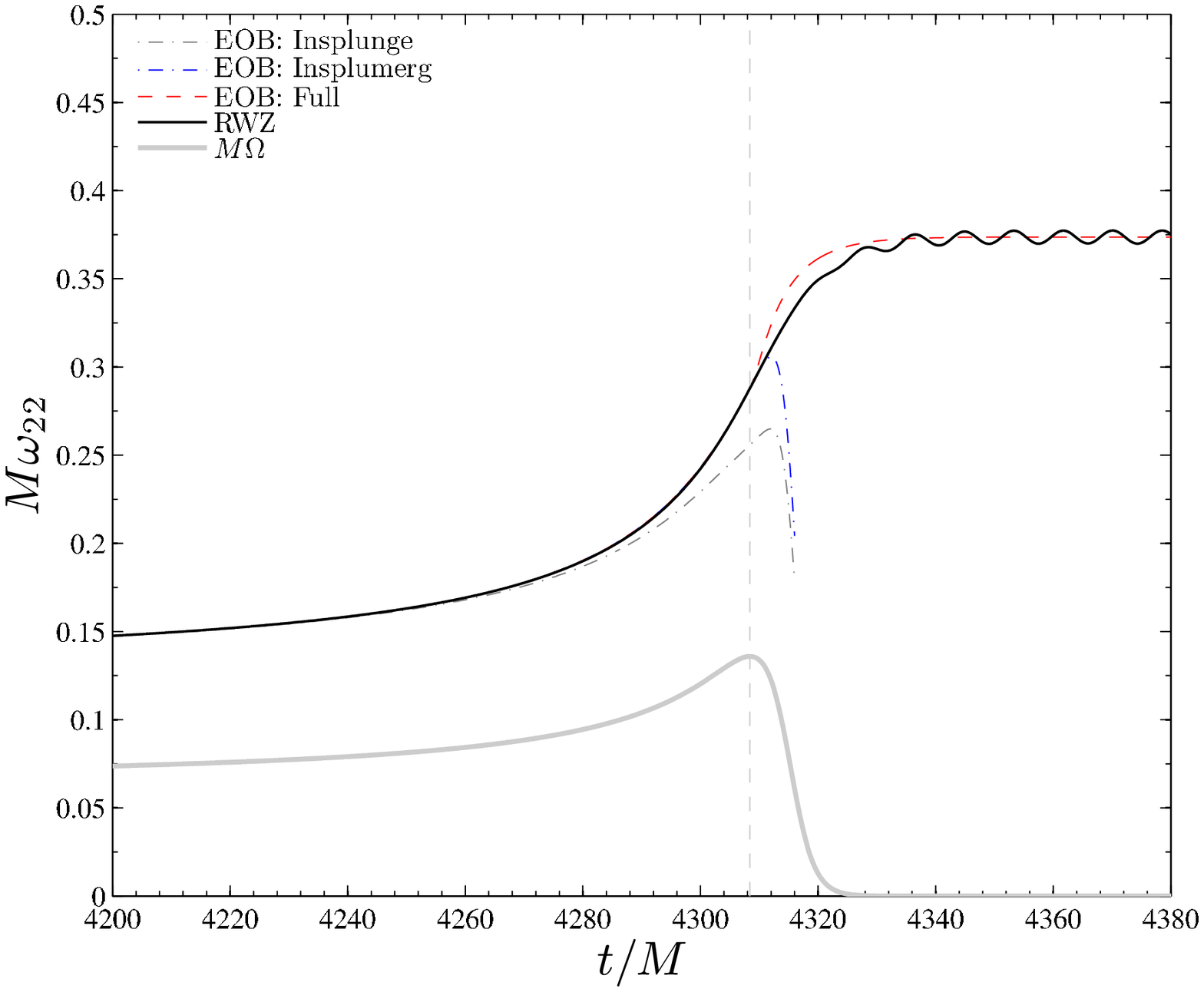}
\hspace{5 mm}
  \includegraphics[width=0.45\textwidth]{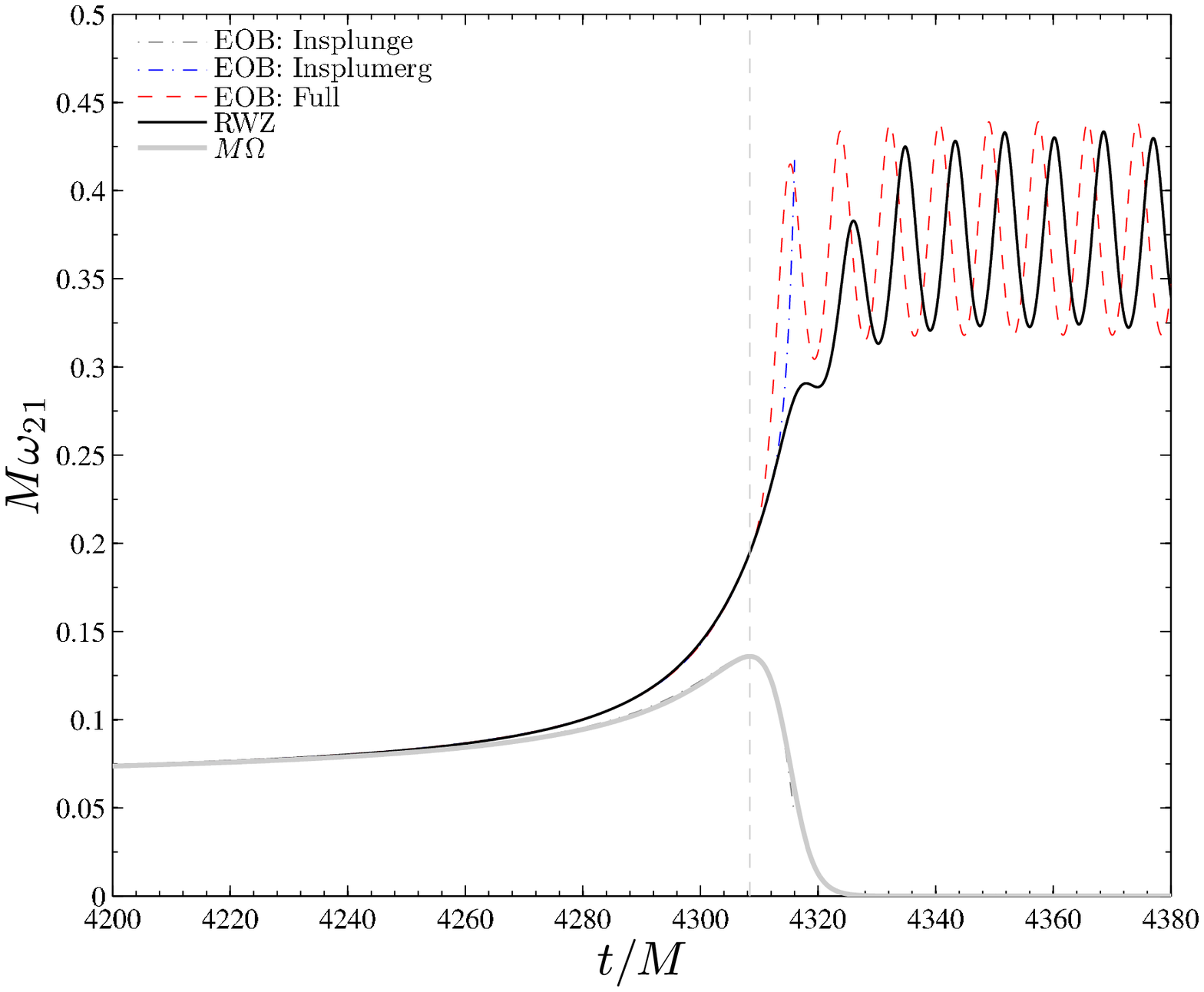}\\
  \caption{\label{fig:l2} Addition of NQC corrections and matching to QNMs; $\ell=2$ multipoles. 
           The (light) dashed lines refer to the bare insplunge waveform, without the addition
           of NQC corrections (dash-dotted line, blue online) nor of QNM ringdown (dashed line, red online). 
           The vertical (light) dashed line indicates the location of the maximum of $M\Omega$.}
\end{center} 
\end{figure*}

\subsection{Compatibility conditions at merger}

Let us focus now on the late-time part of the waveform 
corresponding to plunge, merger and ringdown, \ie $t>t_{\rm LSO}$.
The discussion mainly focusses again on the $\ell=2$ modes as a 
paradigmatic example. The relevant information is collected in 
Fig.~\ref{fig:l2}. 
The top panels show the various waveform moduli
divided by $\nu$, \ie~$A_{2 m}/\nu\equiv |\Psi_{2 m}|/\nu$, with self-explanatory
labelling. The bottom panels show the corresponding instantaneous GW frequencies 
$M\omega_{2 m}=-\Im\left(\dot{\Psi}_{\ell m}/\Psi_{\ell m}\right)$.
For reference also the orbital frequency $M\Omega$ is depicted on all
panels. The vertical dashed line indicates the location of the maximum
of $\Omega$ at $t=t_{\Omega_{\max}}=4308.4M$, that corresponds to the
time $t=t_{\rm LR}$ when the particle crosses the light ring $r_{\rm LR}=3M$.
Note again that we do not allow for any 
arbitrary time or phase shift between EOB and RWZ quantities.  

As previously observed~\cite{Damour:2007xr} the modulus of the insplunge 
$m=2$ waveform (left panel of Fig.~\ref{fig:l2}) is in very 
good agreement during the complete inspiral and during most of the plunge phase 
while overshooting the numerical one by about $10\%$ around $t_{\rm LR}$.
The frequency is indistinguishable by eye up to $t\approx4240$ 
(relative error of $0.5\%$), whereas it clearly underestimates the numerical 
one later on (by $11\%$ at $t_{\rm LR}$). 
Similar results hold for the frequency of the $m=1$ multipole. 
In this case, however, the insplunge amplitude is slightly smaller 
than the numerical one at $t_{\rm LR}$.

\begin{table}[t]
\caption{\label{tab:LR} Strong-field--fast-motion information from the RWZ waveform at the 
light-ring crossing $t=t_{\rm LR}=4308.4M$ and used to determine the $a_i^{\ell m}$ coefficients via
the conditions given by Eqs.~\eqref{eq:cc1}-\eqref{eq:cc4}.}
\begin{center}
\begin{ruledtabular}
\begin{tabular}{llcccc}
$\ell$ & $m$ & $A_\lm(t_{\rm LR})/\nu$ & $A_\lm(t_{\rm LR})/\nu$ & $M\omega_{\ell m}(t_{\rm LR})$& $M\dot{\omega}_\lm(t_{\rm LR})$\\
\hline \hline
2 & 1 & 0.945$\times 10^{-1}$ & 0.271$\times 10^{-2}$ & 0.195$\times 10^{-0}$ & 0.867$\times 10^{-2}$\\
2 & 2 & 0.293$\times 10^{-0}$ & -0.157$\times 10^{-2}$ & 0.288$\times 10^{-0}$ & 0.630$\times 10^{-2}$\\
\hline
3 & 1 & 0.359$\times 10^{-2}$ & 0.238$\times 10^{-3}$ & 0.250$\times 10^{-0}$ & 0.126$\times 10^{-1}$\\
3 & 2 & 0.159$\times 10^{-1}$ & 0.496$\times 10^{-3}$ & 0.350$\times 10^{-0}$ & 0.132$\times 10^{-1}$\\
3 & 3 & 0.514$\times 10^{-1}$ & 0.152$\times 10^{-3}$ & 0.443$\times 10^{-0}$ & 0.106$\times 10^{-1}$\\
\hline
4 & 1 & 0.187$\times 10^{-3}$ & 0.221$\times 10^{-4}$ & 0.321$\times 10^{-0}$ & 0.223$\times 10^{-1}$\\
4 & 2 & 0.118$\times 10^{-2}$ & 0.647$\times 10^{-4}$ & 0.405$\times 10^{-0}$ & 0.191$\times 10^{-1}$\\
4 & 3 & 0.424$\times 10^{-2}$ & 0.152$\times 10^{-3}$ & 0.499$\times 10^{-0}$ & 0.171$\times 10^{-1}$\\
4 & 4 & 0.143$\times 10^{-1}$ & 0.146$\times 10^{-3}$ & 0.592$\times 10^{-0}$ & 0.144$\times 10^{-1}$\\
\end{tabular}
\end{ruledtabular}
\end{center}
\end{table}

As discussed in Sec.~\ref{sec:EOBwaves}, a way to improve the insplunge waveform in the strong-field--fast-motion 
regime is to include the NQC corrections, \ie to consider the insplumerg waveform, 
Eq.~\eqref{eq:insplumerg}.
The insplumerg waveform depends on the $a_i^\lm$ parameters that must be tuned 
requiring some compatibility conditions with the numerical waveform. 
Previous work~\cite{Damour:2007xr,Damour:2009kr,Buonanno:2009qa,Pan:2009wj} 
was mostly restricted to the quadrupolar $m=\ell=2$ waveform and considered only 
NQC amplitude corrections, \ie fixed $a_3^{22}=a_4^{22}=0$ by construction 
in Eq.~\eqref{eq:insplumerg} (see, however, Ref.~\cite{Pan:2009wj} for a preliminary
investigation of the effect of $a_3^{22}$ for spinning binaries).
The amplitude-related parameters in $h^{\rm NQC}_\lm$ were generically fixed by 
imposing that (i) the maximum of the EOB waveform amplitude occurs at the time when the EOB orbital frequency peaks, $t_{\Omega_{\max}}$, and
(ii) the maxima of the EOB and the numerical waveforms amplitude 
agree at $t_{\Omega_{\max}}$. 
When only two parameters are present~\cite{Damour:2009kr}, these conditions 
are sufficient to fix $a_i^\lm$ with $i=1,2$. Additional parameters are tuned by locally 
fitting the numerical waveform~\cite{Buonanno:2009qa,Pan:2009wj}. 
Our procedure builds upon previous works~\cite{Damour:2007xr,Damour:2009kr} with important 
differences. First, we take into account both phase corrections in Eq.~\eqref{eq:fNQC}. The need for NQC phase 
corrections is motivated the frequency plots in Fig.~\ref{fig:l2} (note 
the rather large ``gap'' between the RWZ and the EOB insplunge frequencies at $t_{\rm LR}$)
and requires two more conditions to fix the $a_i^\lm$ parameters with $i=3,4$.
Second, we determine the four parameters $a_i^\lm$ by demanding agreement of 
EOB and RWZ waveforms in amplitude, frequency, and their first derivatives at 
a given time $t=t_m$.
Third, the procedure is applied to all multipoles. 
In formulas, our conditions read
\begin{align}
\label{eq:cc1}
A_\lm^{\rm EOB}(t_m) &= A_\lm^{\rm RWZ}(t_m),\\
\label{eq:cc2}
\dot{A}_\lm^{\rm EOB}(t_m) &= \dot{A}_\lm^{\rm RWZ}(t_m), \\
\label{eq:cc3}
\omega_\lm^{\rm EOB}(t_m) &= \omega_\lm^{\rm RWZ}(t_m),\\
\label{eq:cc4}
\dot{\omega}_\lm^{\rm EOB}(t_m) &= \dot{\omega}_\lm^{\rm RWZ}(t_m) .
\end{align}
The necessary information extracted from the RWZ waveform is displayed 
in Table~\ref{tab:LR}. The only remaining freedom in this procedure is 
the choice of $t_m$. According to the usual EOB prescription we choose 
it from the EOB dynamics as the time $t_{\Omega_{\max}}$ when the orbital
frequency peaks, \ie the time when the small black hole crosses the light ring
at $3M$. In our setting we have $t_m=t_{\rm LR}=t_{\Omega_{\max}}=4308.4M$.
We note that $t_{\rm LR}$ does not exactly coincide with 
the time locations $t^{\lm}_{\max}$ of the peaks of the $A_\lm$'s. 
In particular, $A_{22}$ peaks $2.56M$ {\it earlier} than $t_{\rm LR}$, 
while $A_{21}$ peaks $9.40M$ {\it later}.
Thus the maxima of the amplitude of the multipolar waveform have no strict
relation with the light-ring crossing, but they remain clearly identifiable 
points in the waveforms. An analysis of the amplitude maxima is
reported in Appendix~\ref{app:maxima}. 
The information in Fig.~\ref{fig:l2} is completed by
Table~\ref{tab:maxima} in Appendix~\ref{app:maxima}. 
The third column of the table lists the time shift
$t_{\max}^\lm - t_{\rm LR}$ with respect to the light-ring crossing 
$t_{\rm LR}$ at which each multipolar amplitude peaks.

The $\ell=2$ insplumerg amplitudes and frequencies with NQC parameters determined from 
Eqs.~\eqref{eq:cc1}-\eqref{eq:cc4} are shown as dash-dotted lines (blue online) 
in Fig.~\ref{fig:l2}. The insplumerg is very effective in reproducing 
the numerical data. The corresponding maximum phase difference accumulated 
between $t_{\rm LSO}$ and $t_{\rm LR}$ amounts to $0.025$ rad for both 
multipoles, and the maximum relative difference in amplitude is below 
$2\%$ (see also Figs.~\ref{fig:dphi_lm} and~\ref{fig:dA_lm} below).

The complete EOB amplitude and frequencies with the QNM contribution matched
at $t_{\rm LR}$ are shown as dashed lines (red online) in the four panels 
of Fig.~\ref{fig:l2} (labeled as EOB: Full).
We match both multipoles to three QNMs, using the values
computed in~\cite{Berti:2005ys}.
For $m=2$ we use three positive frequency 
modes, $\sigma^+_{n2}$, with $n=(0,1,2)$ and we neglect the 
negative-frequency mode contribution because it is very 
small~\cite{Bernuzzi:2010ty}. 
For $m=1$ we use two positive frequency modes, $\sigma^+_{n2}$, with $n=(0,1)$ 
and one negative mode $\sigma^-_{02}$, to qualitatively reproduce the related 
oscillation in $M\omega_{21}$. The center of the 3-point matching interval 
$[t_m-\Delta/2,t_m+ \Delta/2]$ (``comb'') is chosen $t_m=t_{\rm LR}$ and 
its width is $\Delta=M$. The matching procedure is robust for 
different choices of $\Delta$ and employs a minimum number of QNMs. 
A larger matching interval~\cite{Damour:2007xr} does not yield any 
improvement in the current setting, while a pointwise matching
($\Delta\to 0$) leads to inaccurate results.
From the figure it is clear that the representation of the ringdown 
does capture the behavior of the numerical waveform more accurately
in the $m=2$  case and less accurately in the $m=1$ case.
This is a consequence of two facts. First, in the EOB
framework the transition from merger to QNM ringdown is 
localized, by construction, at one single point $t=t_m$. 
Second, we choose to determine the NQC parameters and to
match to QNMs at the same time, namely $t_m=t_{\rm LR}$.
We have explored in Appendix~\ref{app:QNM_shift} the
possibility of matching QNMs at a shifted time 
(common to all multipoles) $t_{\rm match}>t_{\rm LR}$ 
while keeping the tuning of NQC corrections fixed at $t_{\rm merger}=t_{\rm LR}$. 
The agreement of modulus, frequency, and phase 
during ringdown significantly improves for all multipoles, 
at the price, however, of one arbitrary shift 
parameter, namely $\Delta t=t_{\rm match}-t_{\rm merger}$.

\begin{table}[t]
  \caption{\label{tab:ai}Values of the NQC coefficients 
  $a_i^\lm$ entering the $h_\lm^{\rm NQC}$ factor, Eq.~\eqref{eq:fNQC}.
  These numbers are obtained imposing the compatibility conditions
  Eqs.~\eqref{eq:cc1}-\eqref{eq:cc4} between EOB and RWZ waveforms at $t=t_m$.}
  \begin{center}
    \begin{ruledtabular}
    \begin{tabular}{llcccc}
      $\ell$ & $m$ & $a_1^\lm$ & $a_2^\lm$ & $a_3^\lm$ & $a_4^\lm$ \\
      \hline 
      2& 1&	  0.0316&	 -0.2874&	  0.7682&	 -0.5872 \\
      2& 2&	  0.0173&	  0.9782&	  0.5019&	 -0.4739 \\
      \hline 
      3& 1&	  2.8720&	 -8.4234&	  1.5859&	 -1.3737 \\
      3& 2&	 -0.0043&	  0.4691&	  1.0368&	 -0.7618 \\
      3& 3&	  0.0722&	  0.9354&	  0.7202&	 -0.5865 \\
      \hline 
      4& 1&	  5.0865&	 -23.6007&	  2.5108&	 -2.0970 \\
      4& 2&	  1.8166&	 -4.5120&	  1.7064&	 -1.2990 \\
      4& 3&	  0.0058&	  0.8158&	  1.1501&	 -0.7671 \\
      4& 4&	  0.1120&	  1.3162&	  0.9295&	 -0.7227 \\
    \end{tabular}
  \end{ruledtabular}
\end{center}
\end{table}

\section{Complete multipolar waveform}
\label{sec:htotal}

In this section we evaluate the performance of the procedure 
discussed above to determine the NQC parameter $a_i$ when it is 
applied to the other multipoles with $\ell>2$.
In the end we put together all the multipolar information to
obtain the complete $\R(h_+-\ii h_\times)$ waveform given by 
Eq.~\eqref{eq:hplus_cross}.

\begin{figure}[t]
\begin{center}
 \includegraphics[width=0.45\textwidth]{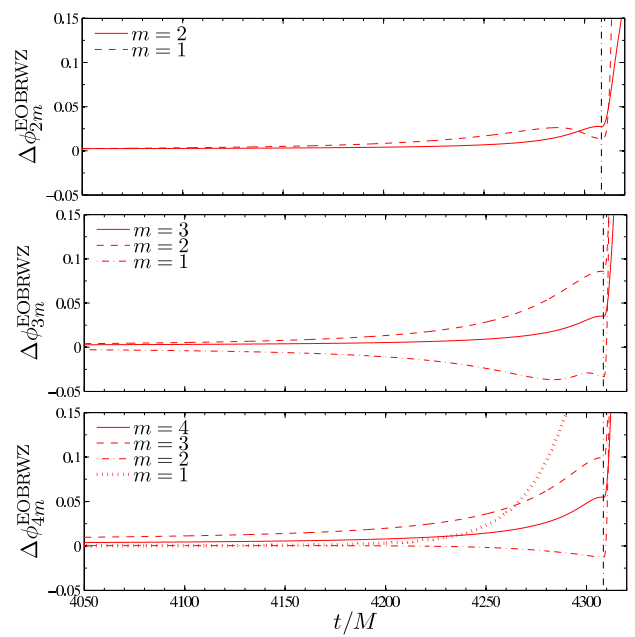}
  \caption{\label{fig:dphi_lm}Time evolution of the phase difference 
   $\Delta\phi^{\rm EOBRWZ}_{\ell m}=\phi^{\rm EOB}_{\ell m}-\phi^{\rm RWZ}_{\ell m}$ 
   between the full EOB and RWZ multipolar waveforms. The dash-dotted vertical line 
   locates the light ring.}
\end{center} 
\end{figure}

\begin{figure}[t]
\begin{center}
  \includegraphics[width=0.45\textwidth]{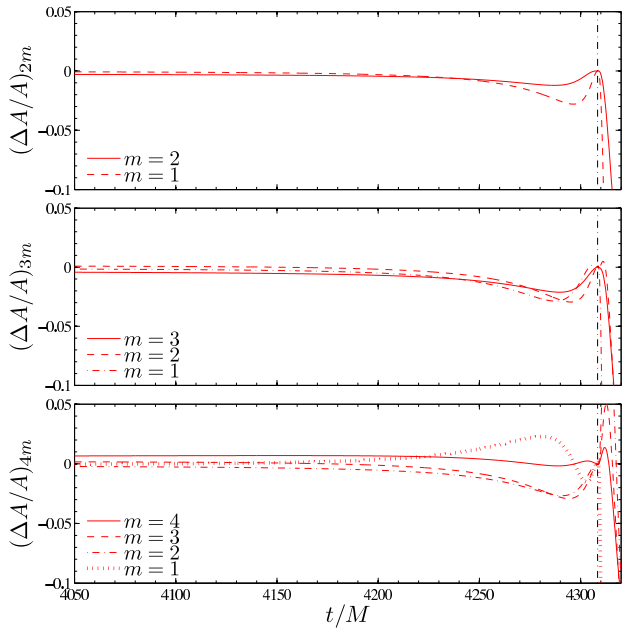}
  \caption{\label{fig:dA_lm}Time evolution of the 
  relative amplitude difference 
  $(\Delta A/A)_\lm = (A_\lm^{\rm EOB}- A_\lm^{\rm RWZ})/A_\lm^{\rm RWZ}$
   between the full EOB and RWZ multipolar waveforms. The dash-dotted vertical 
   line locates the light ring.}
\end{center} 
\end{figure}

\subsection{Multipoles with $\ell >2$}

The qualitative behavior of the insplunge waveform during the 
late-plunge phase for $\ell>2$ is analogous to the $\ell=2$ case. 
The $m=\ell$ EOB waveform amplitude is always slightly
larger than the corresponding numerical one around merger 
and progressively smaller when  $1\leq m <\ell$.  
Analogously, the gap between RWZ and EOB frequency is always smaller
when $m=\ell$ and progressively larger when $m$ decreases.
This suggests that one should obtain numerical values for
the  $a_\lm^i$ coefficients that are systematically larger when 
$m$ decreases, as we actually find (see Table~\ref{tab:ai}).
Note in addition that the value of $a_2^{22}$ is of order
unity, while that of $a_1^{22}$ is of order $10^{-2}$
(as well as most of the others).
These numerical values indicate the consistency of the 
procedure when it is applied to the EMRL case or to 
the comparable mass case.  The following values were obtained
in~\cite{Damour:2009kr}: $a_1^{22}=-0.036\,347$ and $a_2^{22}=1.2468$ in
the equal-mass  
case ($\nu=1/4$) and $a_1^{22}=-0.017\,017$ and $a_2^{22}=1.1906$
in the 2:1 mass case ($\nu=2/9$). The variation of the 
coefficients $a_i^{22}$ is mild when $\nu$ varies between $0$ 
and $1/4$, and it indicates that most of the $\nu$ dependence is already accounted for
by the linear momentum $p_{r_*}$ and its first time derivative. 
It will be interesting  to investigate whether this consistency 
remains (and at what accuracy) for the other multipoles.

Figures~\ref{fig:dphi_lm} and~\ref{fig:dA_lm} quantify the 
phase differences $\Delta\phi^{\rm EOBRWZ}_\lm$ 
as well as the relative amplitude difference 
$(\Delta A/A)_\lm$ for all multipoles. These quantities remain quite
small until the light ring: $\max(\Delta\phi^{\rm EOBRWZ}_\lm)\lesssim
0.15$ rad and $|(\Delta A/A)_\lm|\lesssim 2.5\%$.
Table~\ref{tab:full_LSO} (that is the analogue of Table~\ref{tab:LSO_insplunge})
complements the late-time information given by Figs.~\ref{fig:dphi_lm} and~\ref{fig:dA_lm}
by listing the numerical values of the phase difference and the relative amplitude 
difference at $t_0=500M$ and at the LSO crossing.
The accuracy obtained until the light ring is then lost during the
ringdown part, especially for subdominant multipoles. Because
we are determining the  NQC phase-correction parameters $(a_3^\lm,a_4^\lm)$ by
means of one condition on  $\omega=\dot{\phi}$ and one on $\dot{\omega}_\lm=\ddot{\phi}_\lm$, 
the phase difference itself {\it is not} exactly zero at the 
matching point $t_m=t_{\rm LR}$ (as is the case for the amplitude difference).
This choice of imposing compatibility on $\omega_\lm$ and $\dot{\omega}_\lm$ 
might look overcomplicated: One could just impose compatibility of the
phase and its first time derivative. Working only with
derivatives of the phase may allow us to extend our procedure 
to the comparable mass case using NR data to tune the NQC corrections.
While in this work we have an  unambiguous correspondence between the 
dynamics and the numerical waveform, in the NR simulation this is not the case:
The dynamics is not evidently available and one must typically rely only 
on waveform information that comes with some arbitrary initial phase.
A procedure to fix $(a_3^{\lm},a_4^\lm)$ independently of the GW
phase is then preferable.
In this respect Fig.~\ref{fig:dphi_lm} indicates that compatibility
conditions on $\omega_\lm$ and $\dot{\omega}_\lm$ effectively yield phase 
differences that are quite small at the matching point 
(especially for the $m=\ell$ multipoles). As we shall see in the next 
section, the larger phase differences that are obtained for subdominant 
multipoles at $t_m$ [like the (4,1) case] have no practical influence on the 
complete waveform. 

\begin{table}[t]
  \caption{\label{tab:full_LSO}Phase difference $\Delta\phi^{\rm EOB}\equiv\phi^{\rm EOB}_{\ell m}-\phi^{\rm RWZ}_{\ell m}$ 
   and relative amplitude difference $(\Delta A/A)^{\rm EOB} \equiv (A_\lm^{\rm EOB}- A_\lm^{\rm RWZ})/A_\lm^{\rm RWZ}$ 
   between the full EOB and RWZ waveforms computed at $t_0=500M$ and at the LSO crossing, $t_{\rm LSO}=4076.1M$, 
   for all the ($m\neq 0$) multipoles with $2\leq \ell\leq 4$.}
  \begin{center}
    \begin{ruledtabular}
    \begin{tabular}{llcccc}
      $\ell$ & $m$ & $\Delta\phi_{500M}^{\rm EOB}$ & $\Delta\phi_{\rm LSO}^{\rm EOB}$ & $(\Delta A/A)^{\rm EOB}_{500M}$ & $(\Delta A/A)^{\rm EOB}_{\rm LSO}$ \\
      \hline
2& 1&	 2.49$\times 10^{-4}$ & 	 2.95$\times 10^{-3}$ &	 -1.15$\times 10^{-4}$&	 -8.66$\times 10^{-4}$ \\
2& 2&	 8.79$\times 10^{-4}$ & 	 2.56$\times 10^{-3}$ &	 -1.40$\times 10^{-3}$&	 -3.04$\times 10^{-3}$ \\
      \hline
3& 1&	 -3.3$\times 10^{-4}$ & 	-3.35$\times 10^{-3}$ &	 -5.42$\times 10^{-4}$&	 -1.75$\times 10^{-3}$ \\
3& 2&	 5.18$\times 10^{-4}$ & 	 4.62$\times 10^{-3}$ &	  7.54$\times 10^{-4}$&	  7.98$\times 10^{-4}$ \\
3& 3&	 9.37$\times 10^{-4}$ & 	 2.98$\times 10^{-3}$ &	 -1.98$\times 10^{-3}$&	 -4.40$\times 10^{-3}$ \\
      \hline
4& 1&	 1.19$\times 10^{-4}$ & 	 2.04$\times 10^{-4}$ &	 -2.57$\times 10^{-4}$&	 -1.34$\times 10^{-4}$ \\
4& 2&	 6.6 $\times 10^{-5}$ & 	 3.67$\times 10^{-4}$ &	 -9.48$\times 10^{-4}$&	 -2.45$\times 10^{-3}$ \\
4& 3&	 3.68$\times 10^{-3}$ & 	 1.04$\times 10^{-2}$ &	  1.15$\times 10^{-3}$&	  1.53$\times 10^{-3}$ \\
4& 4&	 1.20$\times 10^{-3}$ & 	 4.12$\times 10^{-3}$ &	  3.41$\times 10^{-3}$&	  6.70$\times 10^{-3}$ 
    \end{tabular}
  \end{ruledtabular}
\end{center}
\end{table}

Finally we tested  the procedure by using as the matching time 
the location of the maximum of the $m=\ell=2$ amplitude
$t_{\max}^{22}$, that occurs  $2.56M$ before $t_{\rm LR}$. 
We measured the corresponding values of the RWZ functions  
and computed the NQC correction factors accordingly. 
Because  $t_{\max}^{22}\simeq t_{\rm LR}$, the EOB waveforms obtained 
in this way are fully compatible with those shown so far.
Moreover, we also explored the effect of using a different matching time for
each multipole, \ie using as $t_m$ the location of the maxima of the various
$A_\lm$. For some multipoles this procedure is very effective in obtaining 
an accurate representation of the modulus, like, for example, in the $(2,1)$
or $(3,1)$ case, although it typically fails to reproduce the frequency. 
The reason for this is that in several situation the matching position 
already corresponds to the final growth of the frequency and the 
simple representation given by our NQC corrections does not 
capture the correct behavior there.
Therefore, we prefer to use the simple procedure discussed so far, 
even if it might lead to a relatively inaccurate representation of the 
ring-down part of the amplitude for certain multipoles 
(as for $\ell=2$, $m=1$).

\subsection{The complete $h_+-\ii h_\times$ waveform}

\begin{figure}[t]
\begin{center}
  \includegraphics[width=0.45\textwidth]{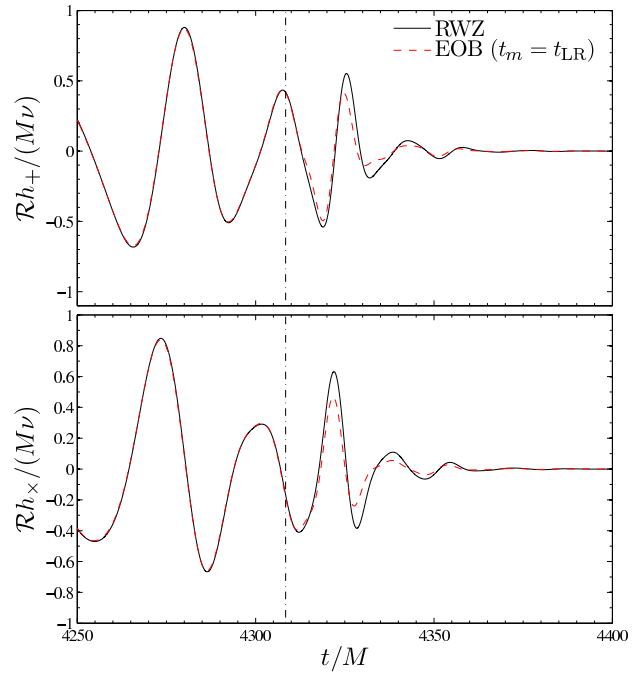}
  \caption{\label{fig:htotal}
    Comparison between the RWZ and EOB complete gravitational waveforms
    taken in the direction $\left(\theta,\Phi\right)=(0,\pi/4)$.
    The two panels show the time evolution of the two polarizations $(h_+,h_\times)$ 
    computed including up to $\ell=4$ multipoles for $\nu=10^{-3}$ 
    (the $m=0$ multipoles are neglected).}
\end{center} 
\end{figure}

\begin{figure}[t]
\begin{center}
  \includegraphics[width=0.45\textwidth]{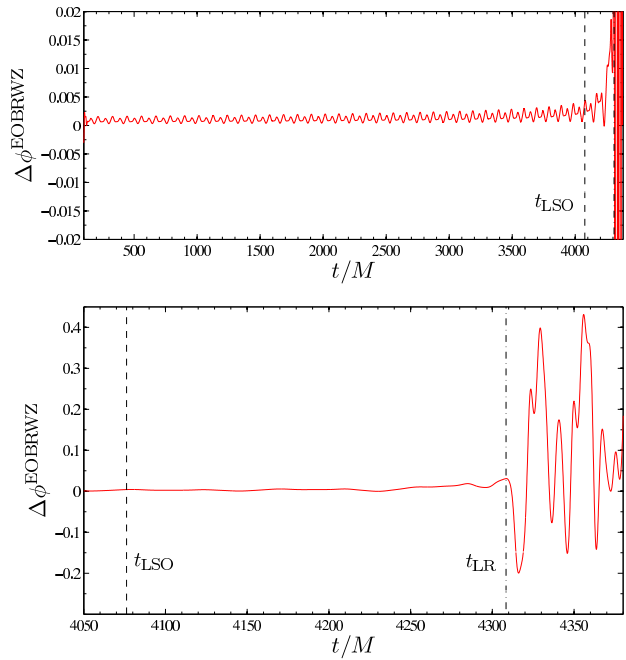}
  \caption{\label{fig:dphihtotal}Time evolution of the phase difference 
   $\Delta\phi^{\rm EOBRWZ}=\phi^{\rm EOB}-\phi^{\rm RWZ}$ between the 
   EOB and RWZ waveforms of Fig.~\ref{fig:htotal}.  The top panel displays 
   the behavior of this function from the very beginning (corresponding to 
   initial separation $r=7M$); the bottom panel focuses on the late-time
   behavior.}
\end{center} 
\end{figure}

Now that we have assessed the quality of the EOB representation of the higher-order
multipoles, let us compare the RWZ and complete EOB gravitational waveforms.
This comparison is shown  in Figs.~\ref{fig:htotal} and~\ref{fig:dphihtotal}.
Figure~\ref{fig:htotal} displays the late-time evolution of the two GW polarizations,
${\cal R}h_+/\nu$ (top) and ${\cal R}h_\times/\nu$ (bottom), that are computed by summing together 
all the $m\neq 0$ multipoles up to $\ell=4$ using Eq.~\eqref{eq:hplus_cross} 
and considering a fiducial direction of emission $\left(\theta,\Phi)=(\pi/4,0\right)$.
The corresponding phase difference is shown in Fig.~\ref{fig:dphihtotal}. 
At the beginning of the inspiral 
(top panel) the dephasing is as small as $1\times 10^{-3}$ rad and is seen 
to progressively increase up to only $4\times 10^{-3}$ rad at the LSO crossing
(dashed vertical line). The phase difference increases then by an order of
magnitude during the plunge phase and reaches about $0.03$ rad at 
the light-ring crossing (dash-dotted vertical line).
The main contribution to the phase difference comes from the $\ell=2$
modes and the $m=\ell$ modes with $\ell>2$. The other modes have a
much smaller 
 impact (especially around merger), even in the EMRL.
The relatively rough description of the ringdown structure (especially for the 
$\ell=2$, $m=1$ mode) implies that the phase difference is larger after the 
light-ring crossing, though it is mainly oscillating around zero.

\section{Conclusions}
\label{sec:conclusions}

In this paper we have discussed the properties of the gravitational 
radiation emitted during the transition from quasicircular inspiral to 
plunge of two nonspinning black holes in the EMRL within the EOB framework.
We considered for the first time the whole multipolar structure of EOB-resummed 
waveforms and we compared and calibrated them against recently calculated 
Regge-Wheeler-Zerilli numerical waveforms~\cite{Bernuzzi:2010ty,BNZ:2010}. 
The target numerical waveforms are extracted at null infinity via the hyperboloidal layer method~\cite{BNZ:2010,Zenginoglu:2007jw,Zenginoglu:2009ey,Zenginoglu:2010cq}.
The binary dynamics is modeled in both cases for a point particle moving on 
a Schwarzschild background under the action of ${\cal O}(\nu)$ dissipative radiation-reaction force
computed using analytically resummed 5PN results~\cite{Damour:2008gu,Fujita:2010xj}.
As a paradigmatic example, we have considered a binary with mass ratio $\nu=10^{-3}$,
initially at a relative separation of $7M$, that inspirals for about 37 orbits before 
plunging into the black hole. The setup of our point-particle ``laboratory'' is 
sufficiently general to allow us to gather information that can be useful both 
in the analytical models of waveforms emitted by EMRI (target sources for 
LISA) as well as for the coalescence of comparable mass black holes (target sources 
for ground-based interferometers). Our results can be summarized as follows.

\emph{Quasiadiabatic inspiral.---}At the beginning of the inspiral, the phase difference between the complete EOB and 
RWZ waveforms (computed without allowing for any relative time and phase shift)
is very small: $\Delta\phi^{\rm EOBRWZ}\sim 5\times 10^{-4}$ rad. This value is consistent with 
the estimated uncertainty related to the residual phases $\delta_{\ell m}$ entering 
the EOB waveform known only up to 4.5PN level.
During the $\sim 33$ orbits of the inspiral after the junk radiation ($t<500M$) 
up to the LSO crossing (corresponding to $\sim 420$ rad of total GW phase) 
the system accumulates only $-2.5\times 10^{-3}$ rad, \ie 
$\Delta\phi^{\rm EOBRWZ}/\phi^{\rm RWZ}=-5.95\times 10^{-6}$.
Such remarkable phase coherence that is obtained with the EOB insplunge 
waveform ---without any tunable parameter--- strongly indicates 
the aptitude of EOB waveforms to model EMRIs for LISA. 

Our conclusions are compatible with those of
Refs.~\cite{Yunes:2009ef,Yunes:2010zj} although there are two important differences.
First, our two waveforms are computed from the same dynamics in order to focus  
{\it only} on waveform comparison so to test the efficiency of the resummation
of the EOB waveform. Second, we do not further calibrate the 
$\nu=0$  EOB-resummed flux (and thus the radiation reaction ${\cal F}_\varphi$) 
to circularized exact data~\cite{Fujita:2009us}. We believe that an additional 
tuning of higher PN contributions to the flux, though necessary for dynamical 
accuracy, would have only a marginal influence on our results.

We finally remark that our setup and the accuracy of our data permitted us to assess the quality of 
the approximation to the $\delta_{\ell m}$'s residual phases entering the EOB waveform. 
Note that we have used the $\delta_{\ell m}$'s in their standard Taylor-expanded 
form~\cite{Damour:2008gu,Fujita:2010xj} and we have not attempted to further
resum them using nonpolynomial expressions. This might certainly be interesting
to explore to further reduce the (small) phase gap we have at the beginning of
the evolution.

\emph{Transition from inspiral to plunge, merger and ringdown.---}For 
the first time we have explored the impact of NQC corrections 
to the complete multipolar waveform, including higher multipoles with $2\leq \ell\leq 4$.
The addition of NQC corrections is important to improve the EOB and RWZ modulus and phase 
agreement towards merger. We have proposed a simple procedure to determine NQC corrections 
on both the phase and amplitude for each multipole;
four parameters are required, two for the amplitude and two for the phase.
They are determined by imposing compatibility between EOB and RWZ waveform amplitude, 
frequency, and their first derivatives at the light ring,
\ie the maximum of the orbital frequency. The procedure is robust 
and applies directly to all multipoles (including those with $\ell>4$, that we have not
explicitly discussed in the text).
The complete EOB gravitational waveform (summed up to $\ell=4$) shows a remarkably 
good phasing and amplitude agreement with the numerical one up to merger ($\pm 0.015$ rad).
After the light ring crossing we have a total phase difference of about $0.25$ due 
to the approximate treatment of the ringdown (via matching to QNMs), although it 
mainly oscillates around zero. The maximum relative amplitude difference 
is about $2.5\%$ just before the light-ring.
We emphasize that the exquisite phase agreement that we find at merger crucially 
relies on the calibration of NQC corrections to the phase.
\begin{figure*}[t]
\begin{center}
  \includegraphics[width=0.45\textwidth]{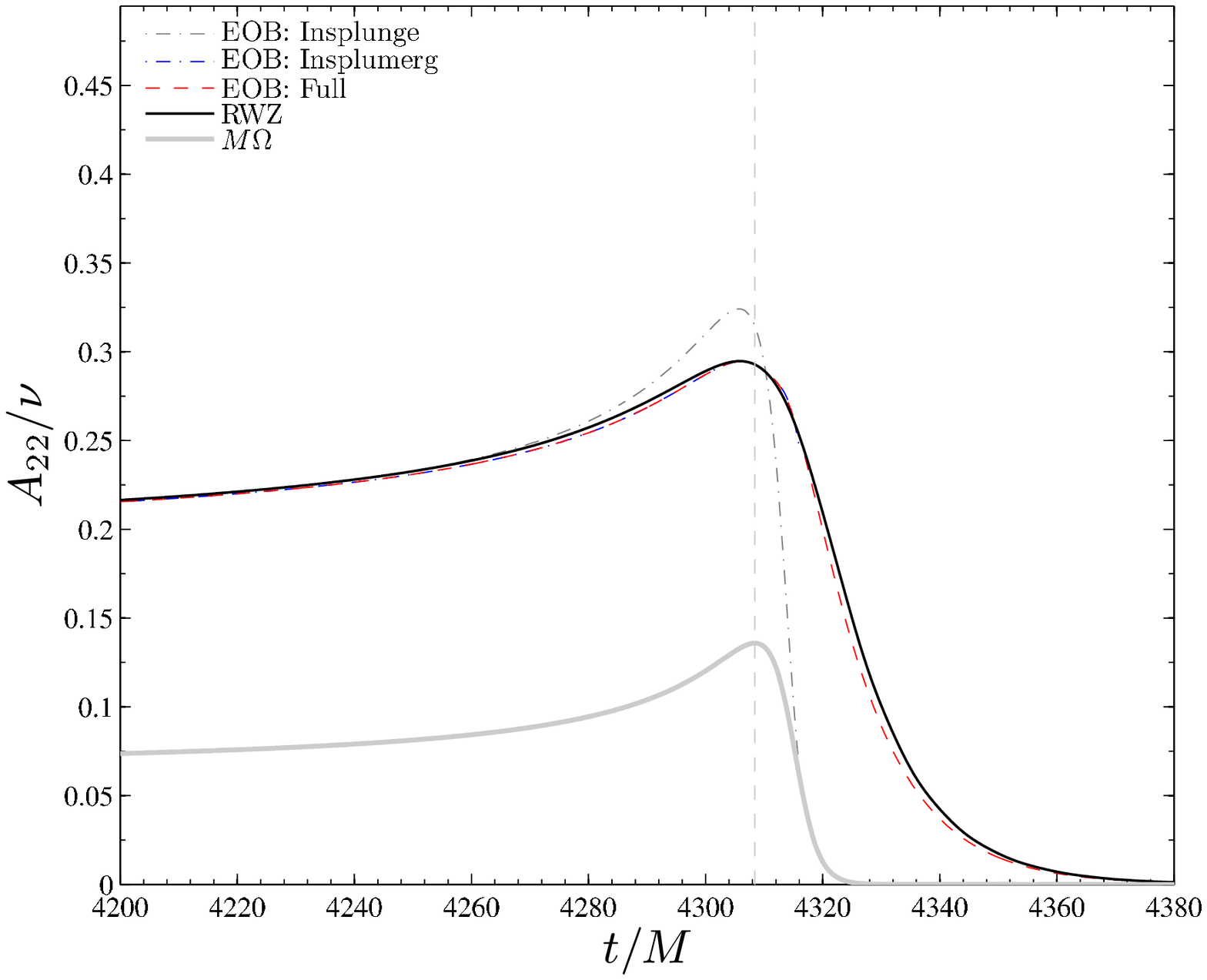}
\hspace{5 mm}
  \includegraphics[width=0.45\textwidth]{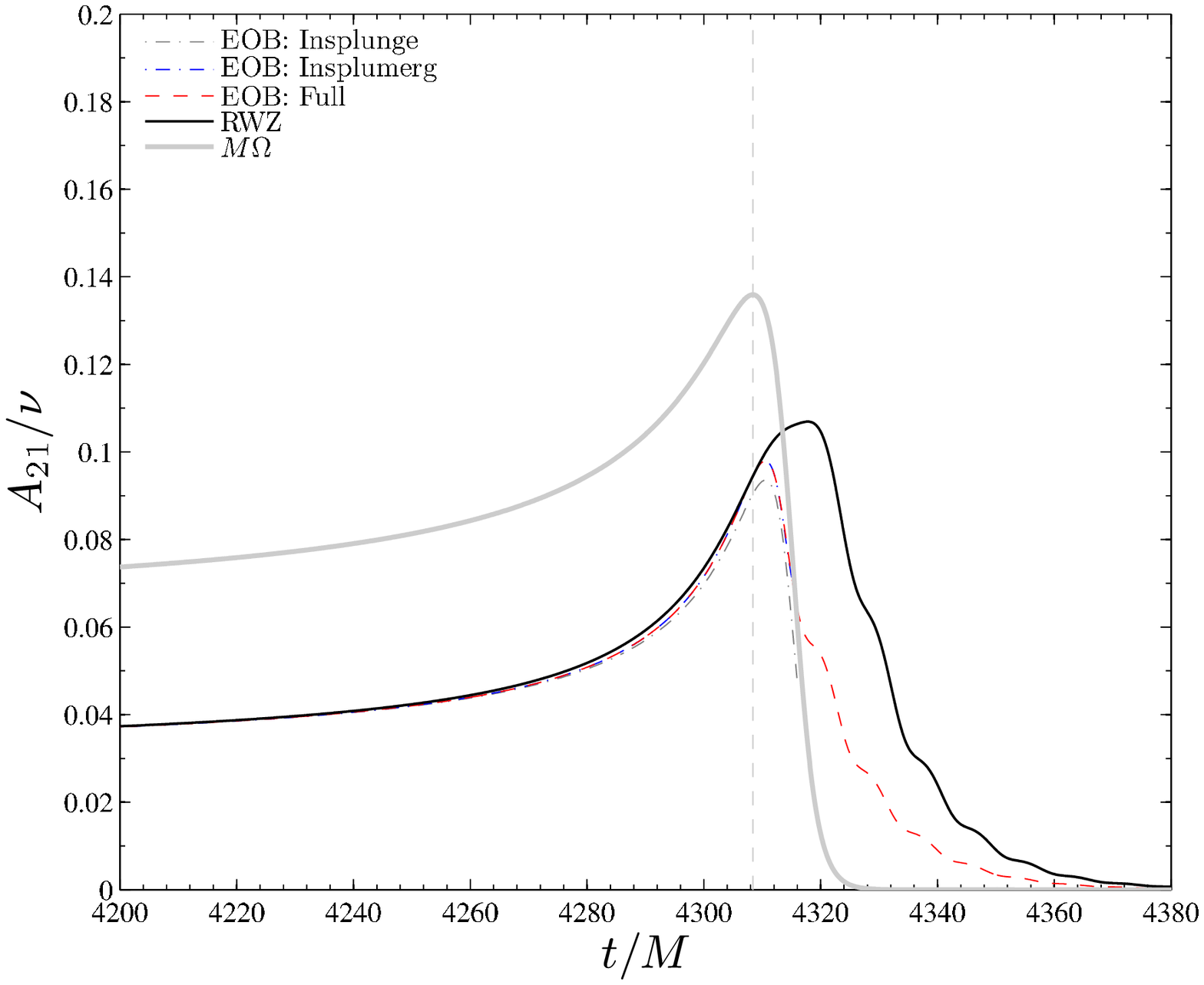}\\
\vspace{5 mm}
  \includegraphics[width=0.45\textwidth]{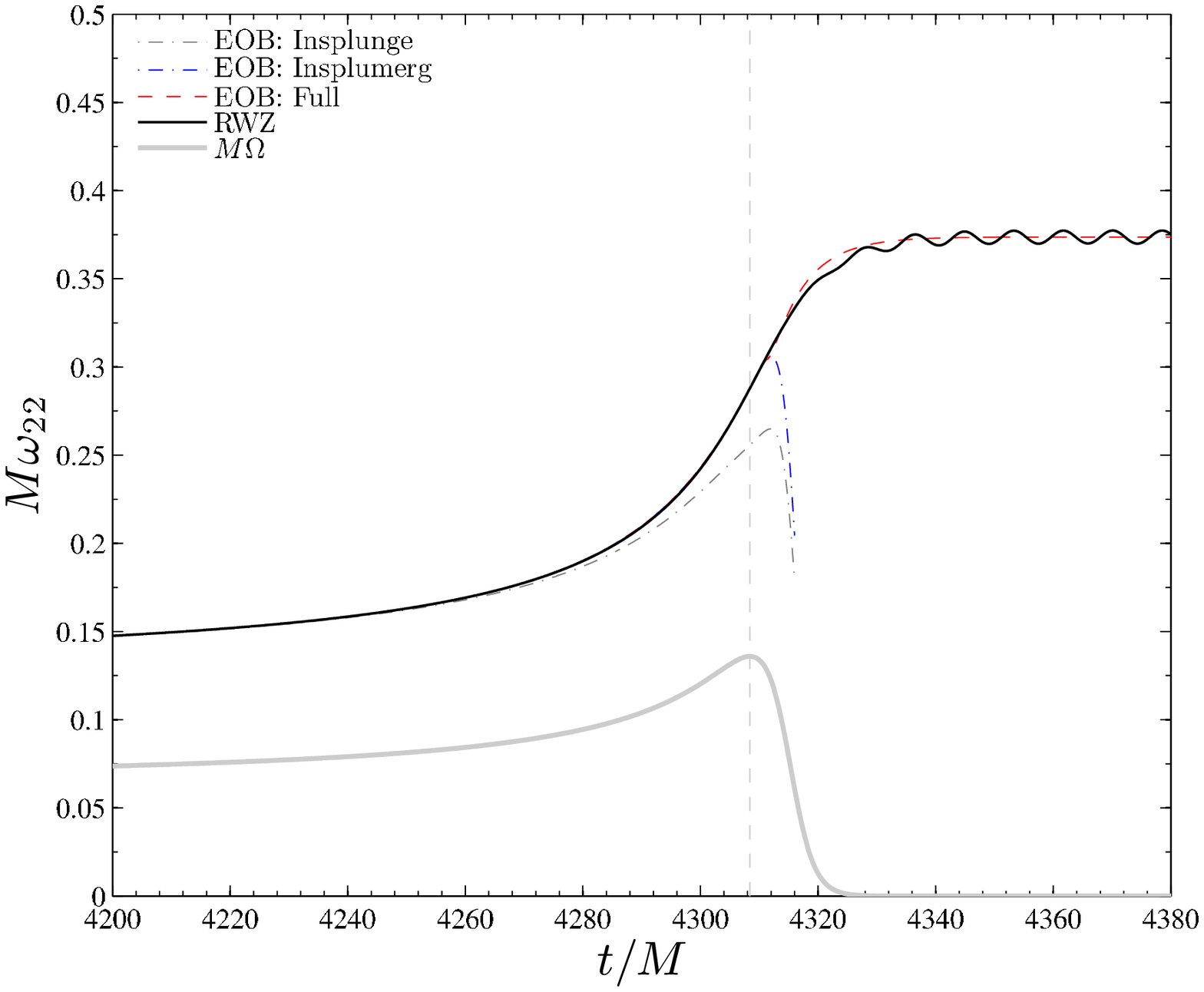}
\hspace{5 mm}
  \includegraphics[width=0.45\textwidth]{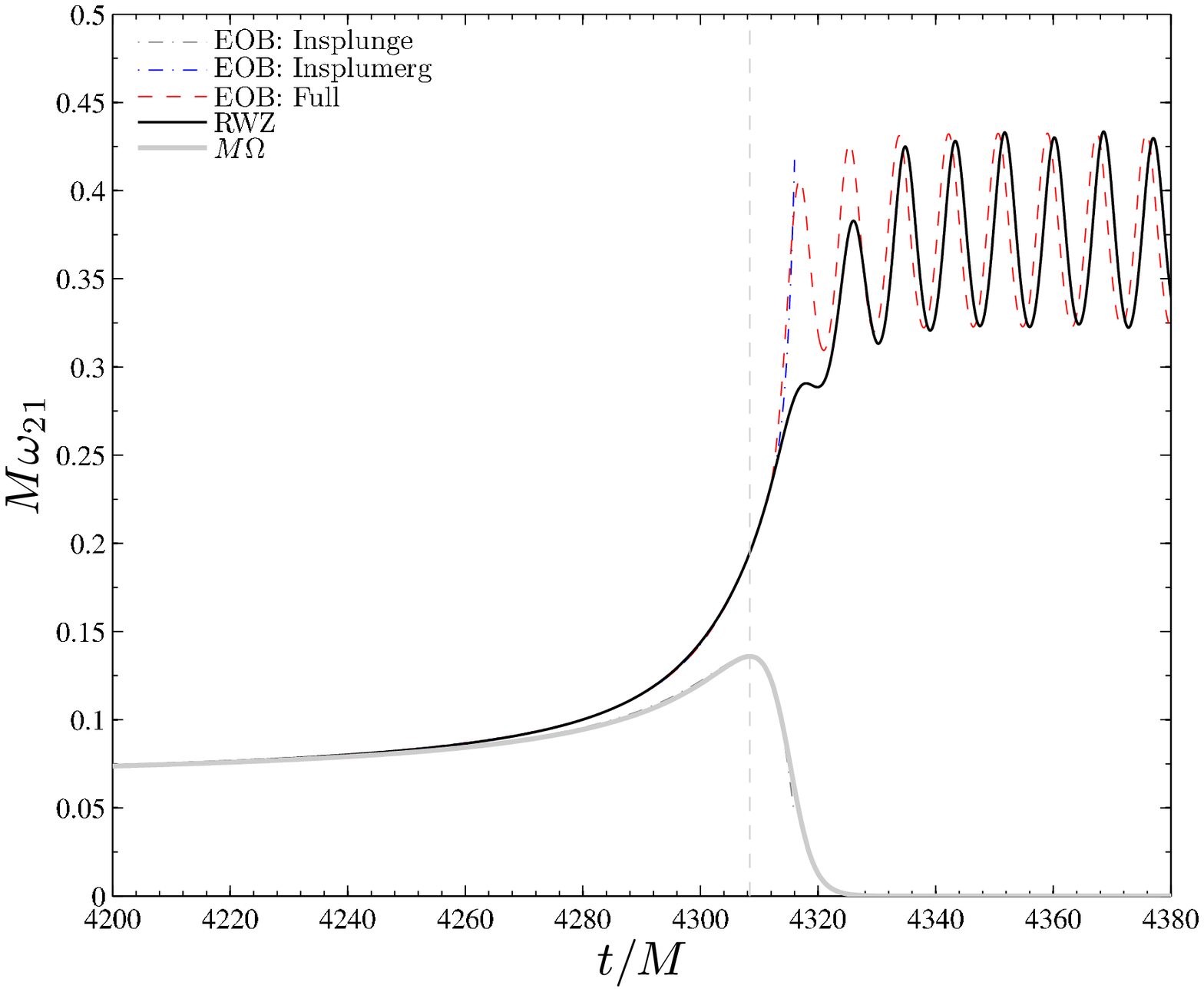}
  \caption{\label{fig:l2shifted} Improvement of the EOB $\ell=2$ ringdown 
   waveform when matching the QNMs at $t_{\rm match}=t_{\rm LR}+3M$. As in Fig.~\ref{fig:l2},
    the (light) dashed lines refer to the bare insplunge waveform, without the addition
    of NQC corrections (dash-dotted line, blue online) nor of QNM ringdown (dashed line, red online). 
    The vertical dashed line locates the maximum of $M\Omega$.}
\end{center} 
\end{figure*}

The procedure discussed here to determine the NQC corrections can be directly applied to 
EOB and NR comparisons for comparable mass ratios  generalizing current techniques. 
However, when $\nu\neq 0$, the procedure is more complicated due to the dependence of 
the EOB dynamics, notably of the Hamiltonian, on other flexibility parameters that 
are also required to be determined (or constrained) by NR data. 
In particular, one of the most evident physical effects entailed by these corrections 
on the dynamics is to displace the location of the ``EOB-light ring'' 
(\ie the maximum of $\Omega$) and thus the location of the matching time $t_m$.
In our $\nu=0$ setting we can identify on the waveform unambiguously the 
time $t_{\rm LR}$ that corresponds to the crossing of 
the light ring, because of the very good ($\sim 10^{-4}$ rad) phase alignment 
of the waveforms at early times and because the underlying dynamics is the same.
This allows us to measure the useful RWZ information at the correct location. 
As we emphasized in the text $t_m$ {\it does not} coincide with the time when $A_{22}$ peaks, 
but it occurs $2.56M$ earlier. 
On the other hand, when dealing with NR data, the exact dynamics is not evidently 
available and one can rely only on waveform information. The peak of the exact
orbital frequency, if it existed, should occur
slightly after the peak of the $A_{22}$ metric waveform amplitude. As a consequence, to apply
the same discussed here to fix the NQC parameters and to keep the maximum of 
$\Omega$ as the matching point, one should measure the four numbers per multipoles slightly 
(say by $\sim 1M$) after the peak of $A_{22}$. This method is different from
 current methods in EOB and NR comparisons, \ie fix the NQC amplitude corrections
imposing that the EOB and NR $A_{22}$ peaks {\it coincide} at the maximum of the EOB orbital frequency.
Even if this procedure is not \emph{a priori} incorrect, we stress that if we were following this
prescription in our setup we would have obtained a  significantly larger phase 
difference ( $\sim~-0.2$ rad), accumulated (starting from the $10^{-2}$ level) in the 
last $50M$ before merger. This suggests that a more detailed analysis of the impact 
of NQC corrections to EOB waveforms in the comparable mass case might be needed in the future.

As a last remark, we emphasize that extraction of numerical waveforms at null infinity 
convinced us to avoid any further (arbitrary) phase and time shift, providing clean 
information about the accuracy of the  analytical modeling of the GW phase in the EOB waveform.
On the contrary, we have shown that waveforms extracted at the finite radius $r_*=1000M$
yield initial phase differences (with the EOB ones) during the inspiral that are $\sim 3\times 10^{-2}$ rad 
for the (2,2) multipole and about twice as much, $\sim 5.5\times 10^{-2}$ rad, for the (2,1) 
multipole (and even larger for the subdominant multipoles~\cite{BNZ:2010}). 
This fact strongly indicates, once more, that in any EOB and NR comparison it 
is {\it necessary} to work either with NR waveforms extrapolated to infinite extraction 
radius or evolved up to null infinity~\cite{Reisswig:2009us,Reisswig:2009rx}.

\acknowledgements

We thank Thibault Damour for useful inputs. SB is supported by DFG Grant 
SFB/Transregio~7 ``Gravitational Wave Astronomy.''  SB thanks IHES for 
hospitality and support during the development of this work.
Computations were performed on the {\tt MERLIN} cluster at IHES. AZ is supported by the Agence Nationale de 
la Recherche (ANR) Grant No.~06-2-134423 in Paris, by the NSF Grant No.~PHY-0601459, 
and by a Sherman Fairchild Foundation grant to Caltech in Pasadena.

\appendix

\section{QNM matching at $t_{\rm match}>t_{\rm LR}$}
\label{app:QNM_shift}

\begin{figure}[t]
\begin{center}
  \includegraphics[width=0.45\textwidth]{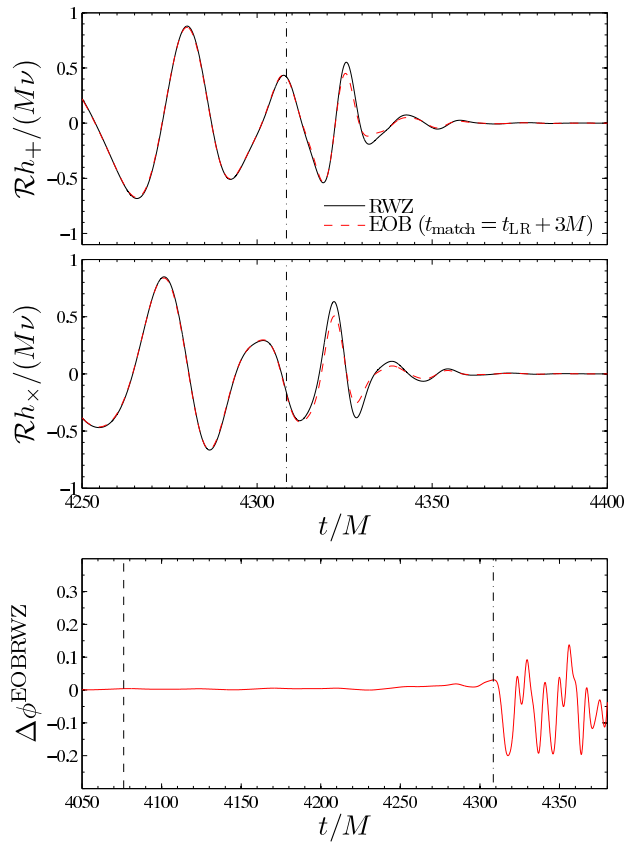}
  \caption{\label{fig:htotal_shift}Improvement of the EOB ringdown waveform
   when matching the QNMs at $t_{\rm match}=t_{\rm LR}+3M$. Top panels: $h_+$ 
   and $h_\times$ GW polarizations. Bottom panel: Phase difference.
   Contrast it with Figs.~\ref{fig:htotal} and~\ref{fig:dphihtotal}.
   The dashed-dotted vertical lines indicates the light-ring crossing.}
\end{center} 
\end{figure}

In this appendix we briefly explore how the quality of the postmerger
waveform depends on shifting the matching to QNMs at  $t_{\rm match}>t_{\Omega_{\max}}=t_{\rm LR}$,
maintaining the determination of the NQC parameters at $t=t_{\rm LR}$.
While the results of Sec.~\ref{sec:compare} and~\ref{sec:htotal} are already satisfactory, 
we experimentally observe that displacing the matching to QNMs by $\Delta t = t_{\rm match}-t_{\rm LR}\sim 3M$ 
produces a further improvement in the final waveform.
This approach is also suggested by a careful inspection of Fig.~\ref{fig:l2} 
in the region around $t_{\rm LR}$: It is clear that (for both multipoles)
the EOB insplumerg frequency and amplitude want to remain close to the numerical one
even {\it after} $t=t_{\rm LR}$, so that we do not seem to be strictly obliged to 
just match the QNMs at $t_m=t_{\rm LR}$. The result of matching QNMs at a shifted 
time is shown in Fig.~\ref{fig:l2shifted}  for $\ell=2$ (compare it with Fig.~\ref{fig:l2}).
The improvement in the closeness between frequencies and amplitude during the
ringdown is striking especially for the $m=2$ mode. An analogous result is obtained
for the other multipoles, the more important improvement affecting the $m=\ell$ modes.
Concerning the $\ell=2$, $m=1$ mode (and similarly for other odd-parity multipoles with $\ell>2$)
one succeeds in having a more accurate representation of the oscillation in the 
frequency due to the interference between positive- and negative-frequency modes.
On the contrary, the amplitude differences in the ringdown are still present.

The influence that the match to QNMs at $t_{\rm match}>t_{\rm LR}$ has on the total 
waveform (summed up to $\ell=4$) can be appreciated in Fig.~\ref{fig:htotal_shift} 
for the two GW polarizations $(h_+,h_\times)$ (top panels; contrast them with Fig.~\ref{fig:htotal})  
and for the total phase difference (bottom panel; contrast it with
Fig.~\ref{fig:dphihtotal}).
Notably, in the latter plot the amplitude of the oscillation in the phase 
difference is more than a factor of 2 smaller than the standard case. 
The result is mainly due to the relatively inaccurate representation of the 
ringdown of the (2,1) mode.

\section{Multipolar hierarchy at merger}
\label{app:maxima}

\begin{figure}[t]
\begin{center}
  \includegraphics[width=0.45\textwidth]{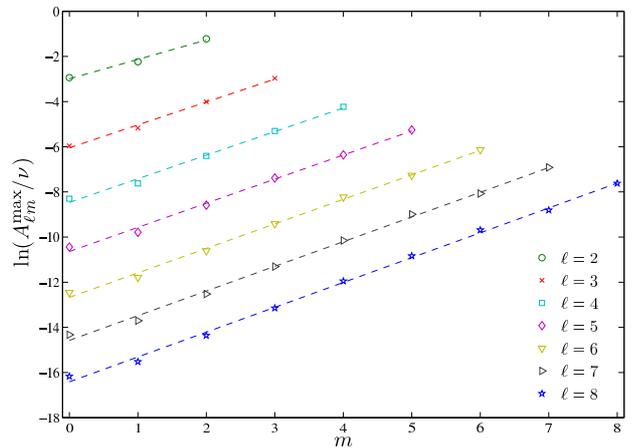}
  \caption{\label{fig:maxima} Maxima of the modulus of the RWZ master
    function for different multipoles. The dashed lines are obtained by
    fitting the data with Eq.~\eqref{eq:fit_max}. }
\end{center} 
\end{figure}

In this appendix we study how the waveform amplitude 
peaks $\Amax=\max(A_{\ell m})$ depend on the multipolar order 
($\ell,m$). The numerical values of $\Amax$ present some 
regularities that it is worth discussing in some detail.
In Fig.~\ref{fig:maxima} we show $\Amax/\nu$ 
for all multipoles up to $\ell=8$. The corresponding numerical 
values are listed in Table~\ref{tab:maxima} for completeness. 
The multipolar structure is such that the peak 
of the waveform amplitude is about 10 times smaller when 
the azimuthal number $\ell$ decreases by one (for
a fixed $m$), while, for a given $\ell$, it increases with $m$. 
As an example, the maximum amplitude of the multipole (4,3)
is larger than the (3,0) one. 

\begin{table*}[t]
  \caption{\label{tab:maxima} Information about the maxima of the modulus of the RWZ multipoles. 
    For each multipole we report %
    the time shift between the time location of the maximum $t_{\max}^\lm$ and the
    light-ring crossing time $t_{\rm LR}$, %
    the maximum of the modulus of the RWZ waveform, its derivative at $t_{\rm LR}$, %
    the frequency, and its derivative at $t_{\max}^\lm$.}
\begin{center}
\begin{ruledtabular}
\begin{tabular}{llcccccccc}
$\ell$&$m$&$(t_{\max}^\lm -t_{\rm LR})$&$\Amax$&$\dot{\Amax}(t^\lm_{\max})$&$M\omega _\lm(t^\lm_{\max})$&$M\dot{\omega}_\lm (t^\lm_{\max})$\\
\hline \hline
2 & 0 & 1.106$\times 10^{1}$ & 0.527$\times 10^{-1}$ & -0.220$\times 10^{-4}$ & 0.790$\times 10^{-21}$& 0.149$\times 10^{-18}$\\
2 & 1 & 9.400$\times 10^{0}$ & 0.107$\times 10^{-0}$ & 0.258$\times 10^{-5}$ & 0.291$\times 10^{-0}$& 0.591$\times 10^{-3}$\\
2 & 2 & -2.560$\times 10^{0}$ & 0.295$\times 10^{-0}$ & -0.260$\times 10^{-5}$ & 0.272$\times 10^{-0}$& 0.587$\times 10^{-2}$\\
\hline
3 & 0 & 1.036$\times 10^{1}$ & 0.258$\times 10^{-2}$ & 0.208$\times 10^{-5}$ & 0.313$\times 10^{-19}$& -0.739$\times 10^{-17}$\\
3 & 1 & 1.054$\times 10^{1}$ & 0.569$\times 10^{-2}$ & 0.767$\times 10^{-6}$ & 0.411$\times 10^{-0}$& 0.679$\times 10^{-2}$\\
3 & 2 & 6.840$\times 10^{0}$ & 0.182$\times 10^{-1}$ & -0.108$\times 10^{-5}$ & 0.452$\times 10^{-0}$& 0.159$\times 10^{-1}$\\
3 & 3 & 1.000$\times 10^{0}$ & 0.515$\times 10^{-1}$ & -0.519$\times 10^{-6}$ & 0.453$\times 10^{-0}$& 0.109$\times 10^{-1}$\\
\hline
4 & 0 & 1.508$\times 10^{1}$ & 0.248$\times 10^{-3}$ & 0.321$\times 10^{-6}$ & -0.523$\times 10^{-18}$& 0.190$\times 10^{-15}$\\
4 & 1 & 1.072$\times 10^{1}$ & 0.488$\times 10^{-3}$ & 0.572$\times 10^{-7}$ & 0.552$\times 10^{-0}$& 0.371$\times 10^{-1}$\\
4 & 2 & 9.520$\times 10^{0}$ & 0.165$\times 10^{-2}$ & 0.158$\times 10^{-7}$ & 0.626$\times 10^{-0}$& 0.196$\times 10^{-1}$\\
4 & 3 & 7.200$\times 10^{0}$ & 0.497$\times 10^{-2}$ & 0.686$\times 10^{-7}$ & 0.637$\times 10^{-0}$& 0.189$\times 10^{-1}$\\
4 & 4 & 2.820$\times 10^{0}$ & 0.145$\times 10^{-1}$ & -0.225$\times 10^{-6}$ & 0.634$\times 10^{-0}$& 0.152$\times 10^{-1}$\\
\hline
5 & 0 & 1.424$\times 10^{1}$ & 0.293$\times 10^{-4}$ & -0.178$\times 10^{-6}$ & -0.208$\times 10^{-16}$& -0.244$\times 10^{-14}$\\
5 & 1 & 1.422$\times 10^{1}$ & 0.562$\times 10^{-4}$ & 0.547$\times 10^{-7}$ & 0.745$\times 10^{-0}$& -0.110$\times 10^{-1}$\\
5 & 2 & 1.122$\times 10^{1}$ & 0.186$\times 10^{-3}$ & 0.390$\times 10^{-7}$ & 0.807$\times 10^{-0}$& 0.312$\times 10^{-1}$\\
5 & 3 & 9.140$\times 10^{0}$ & 0.622$\times 10^{-3}$ & -0.501$\times 10^{-7}$ & 0.801$\times 10^{-0}$& 0.263$\times 10^{-1}$\\
5 & 4 & 7.620$\times 10^{0}$ & 0.173$\times 10^{-2}$ & -0.302$\times 10^{-7}$ & 0.822$\times 10^{-0}$& 0.222$\times 10^{-1}$\\
5 & 5 & 4.120$\times 10^{0}$ & 0.521$\times 10^{-2}$ & -0.268$\times 10^{-6}$ & 0.817$\times 10^{-0}$& 0.189$\times 10^{-1}$\\
\hline
6 & 0 & 1.664$\times 10^{1}$ & 0.388$\times 10^{-5}$ & -0.283$\times 10^{-7}$ & 0.188$\times 10^{-15}$& 0.299$\times 10^{-13}$\\
6 & 1 & 1.376$\times 10^{1}$ & 0.758$\times 10^{-5}$ & -0.595$\times 10^{-8}$ & 0.880$\times 10^{-0}$& 0.332$\times 10^{-1}$\\
6 & 2 & 1.332$\times 10^{1}$ & 0.246$\times 10^{-4}$ & 0.461$\times 10^{-8}$ & 1.021$\times 10^{0}$& 0.102$\times 10^{-1}$\\
6 & 3 & 1.128$\times 10^{1}$ & 0.818$\times 10^{-4}$ & -0.199$\times 10^{-8}$ & 1.001$\times 10^{0}$& 0.297$\times 10^{-1}$\\
6 & 4 & 9.240$\times 10^{0}$ & 0.266$\times 10^{-3}$ & 0.254$\times 10^{-7}$ & 0.985$\times 10^{-0}$& 0.288$\times 10^{-1}$\\
6 & 5 & 8.080$\times 10^{0}$ & 0.697$\times 10^{-3}$ & -0.247$\times 10^{-7}$ & 1.008$\times 10^{0}$& 0.249$\times 10^{-1}$\\
6 & 6 & 5.140$\times 10^{0}$ & 0.216$\times 10^{-2}$ & -0.550$\times 10^{-7}$ & 1.002$\times 10^{0}$& 0.222$\times 10^{-1}$\\
\hline
7 & 0 & 1.588$\times 10^{1}$ & 0.598$\times 10^{-6}$ & -0.332$\times 10^{-8}$ & -0.929$\times 10^{-15}$& -0.256$\times 10^{-12}$\\
7 & 1 & 1.588$\times 10^{1}$ & 0.112$\times 10^{-5}$ & -0.212$\times 10^{-8}$ & 1.063$\times 10^{0}$& -0.878$\times 10^{-2}$\\
7 & 2 & 1.364$\times 10^{1}$ & 0.365$\times 10^{-5}$ & -0.350$\times 10^{-9}$ & 1.186$\times 10^{0}$& 0.416$\times 10^{-1}$\\
7 & 3 & 1.252$\times 10^{1}$ & 0.123$\times 10^{-4}$ & -0.128$\times 10^{-8}$ & 1.189$\times 10^{0}$& 0.330$\times 10^{-1}$\\
7 & 4 & 1.128$\times 10^{1}$ & 0.390$\times 10^{-4}$ & -0.131$\times 10^{-8}$ & 1.186$\times 10^{0}$& 0.319$\times 10^{-1}$\\
7 & 5 & 9.460$\times 10^{0}$ & 0.124$\times 10^{-3}$ & 0.505$\times 10^{-8}$ & 1.171$\times 10^{0}$& 0.311$\times 10^{-1}$\\
7 & 6 & 8.540$\times 10^{0}$ & 0.311$\times 10^{-3}$ & -0.238$\times 10^{-7}$ & 1.195$\times 10^{0}$& 0.273$\times 10^{-1}$\\
7 & 7 & 6.000$\times 10^{0}$ & 0.993$\times 10^{-3}$ & -0.692$\times 10^{-7}$ & 1.188$\times 10^{0}$& 0.250$\times 10^{-1}$\\
\hline
8 & 0 & 1.746$\times 10^{1}$ & 0.951$\times 10^{-7}$ & 0.700$\times 10^{-9}$ & -0.774$\times 10^{-14}$& 0.223$\times 10^{-11}$\\
8 & 1 & 1.534$\times 10^{1}$ & 0.181$\times 10^{-6}$ & -0.336$\times 10^{-9}$ & 1.199$\times 10^{0}$& 0.460$\times 10^{-1}$\\
8 & 2 & 1.524$\times 10^{1}$ & 0.579$\times 10^{-6}$ & 0.141$\times 10^{-9}$ & 1.400$\times 10^{0}$& 0.728$\times 10^{-2}$\\
8 & 3 & 1.366$\times 10^{1}$ & 0.195$\times 10^{-5}$ & 0.205$\times 10^{-9}$ & 1.386$\times 10^{0}$& 0.364$\times 10^{-1}$\\
8 & 4 & 1.242$\times 10^{1}$ & 0.646$\times 10^{-5}$ & -0.699$\times 10^{-9}$ & 1.372$\times 10^{0}$& 0.354$\times 10^{-1}$\\
8 & 5 & 1.136$\times 10^{1}$ & 0.197$\times 10^{-4}$ & 0.547$\times 10^{-10}$ & 1.373$\times 10^{0}$& 0.338$\times 10^{-1}$\\
8 & 6 & 9.720$\times 10^{0}$ & 0.621$\times 10^{-4}$ & 0.396$\times 10^{-8}$ & 1.359$\times 10^{0}$& 0.332$\times 10^{-1}$\\
8 & 7 & 8.960$\times 10^{0}$ & 0.150$\times 10^{-3}$ & 0.153$\times 10^{-7}$ & 1.383$\times 10^{0}$& 0.294$\times 10^{-1}$\\
8 & 8 & 6.720$\times 10^{0}$ & 0.490$\times 10^{-3}$ & 0.238$\times 10^{-7}$ & 1.375$\times 10^{0}$& 0.276$\times 10^{-1}$\\
\end{tabular}
\end{ruledtabular}
\end{center}
\end{table*}

A fit of the $\Amax$ data of Table~\ref{tab:maxima} with the 
functional form
\be
\label{eq:fit_max}
\ln\left(\Amax\right) = c_1(\ell) m + c_2(\ell)\ell
\ee
shows that the behavior of the $\Amax$ is approximately 
exponential in $m$ for each $\ell$; for each value of $m$ $\Amax$, 
scales approximately as $\e^{2\ell}$ when $\ell>2$; 
\ie we have $\Amax\approx \e^{m-2\ell}$. 
The precise numerical values of $c_1(\ell)$ and $c_2(\ell)$ 
are listed in Table~\ref{tab:maxima_fit}, together with
the coefficient of determination $R^2$ that measures
the quality of the fit~\footnote{The coefficient of determination 
  is defined as
  \[R^2\equiv1-\frac{\sum_i(y_i-\hat{y}_i)^2}{\sum_i(y_i-\bar{y})^2}\]
  where $y_i$ is the observed data set, $\bar{y}$ its average 
  and $\hat{y}_i$ the predicted data set. $R^2\in[0,1]$ is the square 
  of the  correlation coefficient~\cite{TaylorBook} and it gives the 
   percent of the variance in $\hat{y}_i$ predictable from $y_i$}. 

\begin{table}[t]
  \caption{\label{tab:maxima_fit} Fit of the maxima of the modulus of
    the RWZ for different $\ell$ multipoles as a function of $m$. 
    The fit assumes the functional form Eq.~\eqref{eq:fit_max}.
    The $90\%$ confidence interval of the estimate coefficients 
    is reported as well as the coefficient of determination $R^2$.}
  \begin{center}
    \begin{ruledtabular}
    \begin{tabular}{lccccc}
      $\ell$ & $c_1(\ell)$ & $90\%$ Conf.& $c_2(\ell)$ & $90\%$ Conf. & $R^2$ \\
      \hline 
	2 &	0.86 & [0.87 0.85]  &	-1.497 & [-1.491 -1.502]	& 0.989 \\
	3 &	1.02 & [1.02 1.00]  &	-2.016 & [-2.013 -2.019]	& 0.995 \\
	4 &	1.05 & [1.06 1.04]  &	-2.116 & [-2.114 -2.119]	& 0.993 \\
	5 &	1.07 & [1.08 1.06]  &	-2.128 & [-2.126 -2.130]	& 0.995 \\
	6 &	1.09 & [1.10 1.08]  &	-2.112 & [-2.111 -2.113]	& 0.997 \\
	7 &	1.09 & [1.10 1.08]  &	-2.082 & [-2.081 -2.083]	& 0.997 \\
	8 &	1.10 & [1.11 1.09]  &	-2.051 & [-2.050 -2.052]        & 0.998 \\
    \end{tabular}
  \end{ruledtabular}
\end{center}
\end{table}

We finally note that the  value for $A^{\rm max}_{22}/\nu$ that
we obtain here is not dramatically different from the values
obtained when the masses $m_1$ and $m_2$ are comparable. 
For example, the value $(A_{22}^{\max}/\nu)_{\nu=0}=0.295$
is about a $10\%$ smaller than the corresponding numerical 
value computed the equal-mass case, $q=m_2/m_1=1$,  
$(A_{22}^{\max})/\nu)_{q=1}\sim 0.321$  and  only about $6\%$ 
smaller when $q=4$, $(A_{22}^{\max}/\nu)_{q=4}\sim 0.31$.
As already pointed out in~\cite{Damour:2009kr}, one can fit
the values of $A^{\rm max}_{22}/\nu$ obtained from a few
(accurate) numerical-relativity simulations as a function
of $\eta=(1-4\nu)$, where $\nu=m_1 m_2/(m_1+m_2)^2$ is the
symmetric mass ratio and reduces to $\nu=\mu/M$ when $m_1\ll m_2$.
Assuming a linear behavior in $\eta$, as it was done 
in~\cite{Damour:2009kr}, one obtains $A_{22}^{\rm max}\simeq 0.321 \nu[ 1 -0.0899(1-4\nu)]$.
However, it turns out that a better representation of
the data is given by a {\it quadratic} behavior 
in $(1-4\nu)$ of the form 
$A_{22}^{\rm max}\simeq 0.321 \nu[ 1 -0.0162(1-4\nu) + 0.0792(1-4\nu)^2]$.
In the future it will be interesting to check the numerical accuracy
of this relation versus other NR data, as well as to generalize it
to the other multipoles.



\end{document}